\documentclass[prx,twocolumn,superscriptaddress,amssymb]{revtex4-1}
\usepackage[usenames,dvipsnames]{color}
\usepackage{psfrag,graphicx}
\usepackage{dcolumn}
\usepackage{bm}
\usepackage{amsfonts,amssymb,amsmath}
\usepackage{tabularx, booktabs}
\usepackage{slashed}
\usepackage{makecell}
\usepackage{diagbox}
\usepackage[colorlinks]{hyperref}

\hypersetup{
plainpages=true,
breaklinks=true,
hypertexnames=false,
 pageanchor=true,
colorlinks=true,
linkcolor={blue},
citecolor={magenta},
urlcolor={blue},
anchorcolor={black}
}

\newcommand{\beq}{\begin{eqnarray}}
\newcommand{\eeq}{\end{eqnarray}}

\newcommand{\bra}[1]{\langle #1|}
\newcommand{\ket}[1]{|#1\rangle}

\newcommand{\expec}[1]{\left\langle #1 \right\rangle}

\newcommand{\figref}[1]{\mbox{Fig.~\ref{#1}}}

\newcommand{\appref}[1]{\mbox{Appendix~\ref{#1}}}
\renewcommand{\eqref}[1]{\mbox{Eq.~(\ref{#1})}}

\newcommand{\be}{\begin{equation}}
\newcommand{\ee}{\end{equation}}
\newcommand{\bea}{\begin{eqnarray}}
\newcommand{\eea}{\end{eqnarray}}

\newenvironment{colored_text}
  {\color{black}}
  {}

\newcommand{\bc}{\begin{colored_text}}
\newcommand{\ec}{\end{colored_text}}

\begin{document}
\title{Virtual excitations in the ultra-strongly-coupled spin-boson model: physical results from unphysical modes}
\date{\today}
\author{Neill Lambert}
\thanks{These authors contributed equally to this work.}
\affiliation{Theoretical Quantum Physics Laboratory, RIKEN Cluster for Pioneering Research, Wako-shi, Saitama 351-0198, Japan}

\author{Shahnawaz Ahmed}
\thanks{These authors contributed equally to this work.}
\affiliation{Theoretical Quantum Physics Laboratory, RIKEN Cluster for Pioneering Research, Wako-shi, Saitama 351-0198, Japan}
\affiliation{BITS Pilani, Goa campus, Sancoale, Goa 403726, India}
\affiliation{Wallenberg Centre for Quantum Technology, Department of Microtechnology and Nanoscience, Chalmers University of Technology, 412 96 Gothenburg, Sweden}

\author{Mauro Cirio}
\affiliation{Graduate School of China Academy of Engineering Physics, No.\;10 Xibeiwang East Road, Haidian District, Beijing, 100193, China}
\affiliation{Theoretical Quantum Physics Laboratory, RIKEN Cluster for Pioneering Research, Wako-shi, Saitama 351-0198, Japan}

\author{Franco Nori}
\affiliation{Theoretical Quantum Physics Laboratory, RIKEN Cluster for Pioneering Research, Wako-shi, Saitama 351-0198, Japan}
\affiliation{Department of Physics, University of Michigan, Ann Arbor, Michigan 48109-1040, USA}

\begin{abstract}

Here we show how, in the ultra-strongly-coupled spin-boson model, apparently unphysical ``Matsubara modes'' are required not only to regulate detailed balance, but also to arrive at a correct and physical description of the non-perturbative dynamics and steady-state. In particular, in the zero-temperature limit, we show that neglecting the Matsubara modes results in an erroneous emission of ``virtual" photons from the collective ground state. To explore this difficult-to-model regime we start by using a non-perturbative hierarchical equations of motion (HEOM) approach,  based on a partial fitting of the bath correlation-function which takes into account the infinite sum of Matsubara frequencies using only a biexponential function.  We compare the HEOM method to both a pseudo-mode model, and the reaction coordinate (RC) mapping, which help explain the nature of the aberrations observed when Matsubara frequencies are neglected.  For the pseudo-mode method we present a general proof of validity, which allows for negative Matsubara-contributions to the decomposition of the bath correlation functions to be described by zero-frequency Matsubara-modes with non-Hermitian coupling to the system. The latter obey a non-Hermitian {\em pseudo}-Schr\"{o}dinger equation, ultimately justifying why superficially unphysical modes can give rise to physical system behavior.
\end{abstract}
\maketitle

\section{Introduction}
\label{intro}
The spin-boson model is a cornerstone of the theory of open-quantum systems, and its elegance often belies its power to describe a wide range of phenomena \cite{leggett1987dynamics, breuer2002theory, ingold2002path}. It not only allows us to understand the relationship between quantum dissipation and classical friction, but is a powerful  model to study  topics ranging from physical chemistry to quantum information. Practically speaking,  a number of perturbative approaches and assumptions such as the Born-Markov and rotating-wave approximation (RWA) are usually employed to obtain tractable solutions.  However, research areas such as energy transport in photosynthetic systems \cite{Akihito09,panitchayangkoon2011direct, Peter11, Lambert_2013,Chen2015,ishizakireview,Scholes2017}, quantum thermodynamics \cite{Strasberg_2016,newman2017performance}, and the ultrastrong coupling regime in artificial light-matter systems  \cite{anappara2009signatures,todorov2010ultrastrong, niemczyk2010circuit, PhysRevLett.105.196402, scalari,Garcia-Ripoll2015,Kockum18b,forn2018ultrastrong},  have demanded the development of numerically exact methods to explore non-perturbative and non-Markovian parameter regimes \cite{de2018dynamics, magazzu2018probing,zhangprl}, which are out of reach of traditional approaches.

In the limit of a discrete single bosonic mode, as arises in circuit QED \cite{Gu17}, the non-perturbative limit, when the coupling is a significant fraction of the cavity frequency, is sometimes referred to as the ultra-strong coupling (USC) regime \cite{Kockum18b,forn2018ultrastrong}. This regime harbours a range of new physics, including higher-order coupling effects, the possibility to excite two systems with one photon \cite{PhysRevLett.117.043601},  preparing Bell and GHZ states in cQED  \cite{marci18} and virtual excitations \cite{anappara2009signatures, Stassi13, DeLIberato17,Kockum17}. In the latter, the excitations are called virtual because they are energetically trapped in the hybridized  light-matter  ground-state.  A correct theoretical understanding of this trapping, such that unphysical emission from the ground-state is avoided, was only developed recently \cite{simoneExtra, simoneComment}.  It is now understood that non-adiabatic external forces must be applied to transmute them into real, observable, excitations \cite{Johansson09,Johansson13b,Stassi13,Cirio16, Cirio18}. Numerical simulations \cite{Zueco2013, Munoz18, zueco2018} have suggested that a similar phenomenon occurs in continuum systems describable with the spin-boson model \cite{DeLIberato17}, like one-dimensional transmission lines \cite{magazzu2018probing, gustafsson2014propagating}, and superconducting metamaterials \cite{scarlino1,scarlino2,scarlino3,scarlino4}.  In addition, at non-zero temperatures it has been shown that these virtual excitations can influence other processes, like the efficiency of a quantum heat-engine \cite{melina}.

To explore these features in continuum systems, we need non-perturbative, and non-Markovian methods, one example of which is the hierarchical equations of motion (HEOM) technique \cite{tanimura1989time, ishizaki2005quantum}. However, traditionally low-temperature regimes are difficult \cite{tang2015extended,fruchtman2016perturbative, duan2017zero}, if not inaccessible, with the HEOM.  This is because the HEOM relies on a decomposition of the bath correlation function into a sum of exponentials.  Unfortunately, due to the physical constraint disallowing Hamiltonians unbound from below (i.e., that the environment only consists of positive frequency modes), even a simple Lorentzian spectral density gives correlation functions which cannot be analytically decomposed into a finite sum.

To overcome this difficulty we separate the correlation function into an analytical part, comprised of a {\em finite} number of exponentials, and the ``Matsubara'' part, given by an {\em infinite} sum of exponentials (the latter of which was neglected in other works studying the zero-temperature limit of the HEOM method \cite{Ma_2012,LZI}). In the zero-temperature limit, we analytically integrate the infinite sum and then fit it with a biexponential function.  Fitting the total correlation-function to exponentials for use with the HEOM has also been explored in \cite{dattani2012optimal, tang2015extended,fruchtman2016perturbative, duan2017zero} but our approach allows us to limit the fitting error \cite{PlenioErrorsCorr} to the Matsubara component, and gives us physical insight into the role of the different contributions to the correlation function. The fitting inevitably introduces some error in the system dynamics, which we analyze in the detail in the appendix.

By comparing results with and without this Matsubara contribution, we find that the neglect of the Matsubara terms in the HEOM formulation induces a very specific error in the dynamics and steady-state. This error corresponds to an unphysical system temperature, even at weak-coupling, due to violation of detailed balance, and the production/emission of unphysical photons  from the ground state of the coupled light-matter system  in the ultra-strong-coupling regime.

To show these features more directly, we compare the HEOM results to both a pseudo-mode model \cite{Garroway} and a reaction coordinate (RC) model \cite{Garg_1985, Martinazzo_2011,iles2014environmental, Strasberg_2016}. For the RC model, we find qualitative agreement to the full HEOM solution for narrow baths, and a clear description of how the Matsubara frequencies are important for trapping excitations in the USC regime: {\em ignoring the Matsubara frequencies is found to be equivalent to making both a rotating-wave and Markov approximation for the interaction between the RC mode and the residual environment}. The latter, directly leads to the unphysical emission of energy from a collective ground-state in the RC picture.

In addition, we find that the pseudo-mode model, also employing the fit of the Matsubara parameters in the form of two additional zero-frequency ``Matsubara modes'' with  non-Hermitian coupling to the system, {\em can exactly reproduce the full HEOM results for all parameter regimes.}  It can also be used to give meaning to the auxiliary density operators (ADOs) of the HEOM, indicating a strong relationship between the two methods.  To account for the unusual form of the ``Matsubara modes'' we explicitly generalize the proof of validity of the pseudo-mode method \cite{Garroway,plenio2018}. Our derivation shows that by combining the \emph{non-Hermitian} Hamiltonian together with what we call a \emph{pseudo}-Schr\"{o}dinger equation, the Dyson equation for the reduced dynamics of the system is formally equivalent to one where the system is \emph{physically} interacting with the original continuum environment.

We begin with an introduction to the spin-boson model and bath-correlation functions. We then provide an intuitive explanation of why omitting the apparently negligible Matsubara terms can have large consequences, even in the weak-coupling regime. We then demonstrate our correlation function fitting method for the HEOM, before turning to the pseudo-mode method and the reaction coordinate mapping to more transparently explain what happens when Matsubara terms are ignored in the ultra-strong coupling regime. Finally, we compare all three methods, with and without Matsubara contributions, and show their predictions for the dynamics and steady-state occupation of certain environment modes.

\section{The spin-boson model}
The iconic spin-boson model considers a two-level system (the spin, or qubit) in a bath of harmonic oscillators with the total system-bath Hamiltonian given by (setting $\hbar=1$ throughout):
\begin{equation}
\label{spinh}
H = \frac{\omega_q}{2}\sigma_z + \frac{\Delta}{2}  \sigma_x + \sum_k \omega_k b_k^{\dagger}b_k + \sigma_z \tilde{X}\;\;,
\end{equation}
where $\omega_q$ is the qubit splitting, $\omega_k$ is the frequency of the $k^{\mathrm{th}}$ bath mode, $\Delta$ is the tunnelling matrix element, $\sigma_{z(x)}$ are the Pauli matrices acting on the qubit. For later use we define  $\bar{\omega}=(\omega_q^2+\Delta^2)^{1/2}/2$, as the free qubit eigenfrequency. The  $k^{\mathrm{th}}$ mode of the bath, associated with annihilation operators $b_k$, interacts with the qubit via the operators $\tilde{X}_k=g_k/\sqrt{2\omega_k}(b_k+b^\dagger_k)$ in terms of the couplings $g_k$, so that $\tilde{X}=\sum_k \tilde{X}_k$.

The effect of the bath can be considerably simplified when the initial state of the environmental modes is Gaussian, and in a product state with the system (the qubit). Specifically, we assume the bath to be in a thermal state at a temperature $T$. In this case the influence of the environment is contained in the two-time correlation function $C(t)=\langle \tilde{X}(t)\tilde{X}(0) \rangle$. The correlation function of the free bath, when it is not in contact with the system, can be written (in the continuum limit) as,
\beq
\label{corrspecial}
C(t) = \frac{1}{\pi} \int_0^{\infty}\!\!\!\!\! d\omega J(\omega) \left[\coth\left(\frac{\beta \omega}{2}\right) \cos(\omega t) - i\sin(\omega t)\right].
\eeq
Here $J(\omega) = \pi \sum_k {g_k^2}/{2\omega_k} \delta(\omega-\omega_k)$ is the spectral density which parameterizes the coupling coefficients $g_k$, and $\beta = {1}/{k_{\text{B}}T}$ is the inverse temperature. Throughout this article we focus on the following  ``underdamped Brownian motion spectral density'',
\beq
\label{underdamp}
J(\omega) &=&\frac{ \gamma \lambda^2\omega}{(\omega^2-\omega_0^2)^2+\gamma^2 \omega^2}\;\;,
\eeq
which is characterized by a resonance frequency $\omega_0$,  a width $\gamma$, and a strength $\lambda$. A spectral density of this form is a convenient basis in which one can represent a range of other spectral densities \cite{tannor-meier, Kreisbeck2012}. \par
\bc In the under-damped limit ($\gamma < 2\omega_0$), \ec it is convenient to decompose the correlation function, for \eqref{underdamp} in \eqref{corrspecial}, as $C(t) = C_0(t) + M(t)$, where
\begin{equation}
\label{eq:nonmatsubara}
\begin{array}{lll}
C_0(t) &=&  \displaystyle\frac{\lambda^2e^{-\gamma t/2}}{4 \Omega}\left[C_0^R(t)+ C^I_0(t)\right]\;\;,
\end{array}
\end{equation}
in terms of $C_0^R=\coth\left[ {\beta(\Omega+i\Gamma)}/{2}\right]\exp{(i\Omega t)}+\text{H.c.}$ (where $\text{H.c.}$ denotes Hermitian conjugation) and $C_0^I=e^{-i\Omega t}-e^{i\Omega t}$, and
\begin{equation}
\label{matsubaracorr}
M(t) = -\frac{2\lambda^2 \gamma}{\beta} \sum_{k > 0}^{\infty} \frac{\omega_k e^{-\omega_k t}}{\left[ (\Omega+i\Gamma)^2 + \omega_k^2\right] \left[ (\Omega-i\Gamma)^2  + \omega_k^2 \right]},
\end{equation}
 with the definitions $\Gamma = \gamma/2$, $\Omega^2 = \omega_0^2-\Gamma^2$, and $\omega_k=2\pi k /\beta$ ($k\in \mathbb{N}$) for the Matsubara frequencies.

Intuitively, the $C_0(t)$ part of the correlation function characterizes the resonant part of the bath, with a shifted resonant frequency $\Omega$ and  decay rate $\gamma/2$. On the other hand, the $M(t)$ part of the correlation function seems to have a less transparent description: it has no resonances but  infinite sub-contributions which decay at rates equal to $\omega_k$ (hence we will name it the ``Matsubara correlation''). One way to explore its meaning is to study what happens to the qubit dynamics after imposing $C(t)\rightarrow C_0(t)$, i.e., completely neglecting it.
Note that this will induce an error even at zero temperature ($\beta\rightarrow\infty$)
due to the competition between the factor $\beta^{-1}$ and the Matsubara frequencies approaching the continuum.

To proceed with our intuitive analysis, it is worth considering the Fourier transform of the correlation function, i.e., the power-spectrum $S(\omega)=\int_{-\infty}^\infty dt~C(t) e^{i\omega t}=J(\omega)[1+\coth(\beta\omega/2)]$. From this expression it is possible to check that the power-spectrum encodes the symmetry condition
\begin{equation}
\label{eq:symm}
{S(\omega)}{}=\exp{(\beta\omega)}S(-\omega)\;\;.
\end{equation}
When the coupling to the environmental degrees of freedom is small compared to the qubit eigenfrequency $\bar{\omega}=(\omega_q^2+\Delta^2)^{1/2}/2$, the effect of the bath can be studied perturbatively (for example by using the Fermi golden rule). In this case, the qubit will absorb (relax)  energy from (into) the environment at rates proportional to $S(-\bar{\omega})$ ($S(\bar{\omega})$) so that Eq.~(\ref{eq:symm}) encodes the physical meaning of the detailed balance condition. As a consequence, by neglecting the Matsubara correlations, we are then going to break this balance \cite{hanggi,majewski1984detailed}. Nevertheless, the qubit will still reach an equilibrium thermal state at the effective temperature
\begin{equation}
\label{beffeq}
\beta_\text{eff}=\displaystyle\frac{1}{\bar{\omega}}\log{\frac{S_0(\bar{\omega})}{S_0(-\bar{\omega})}}\;\;,
\end{equation}
where  $S_0(\omega)=\int_{-\infty}^\infty dt~C_0(t) e^{i\omega t}$. The relation between $\beta_\text{eff}$ and the actual temperature $\beta$ intuitively quantifies the effect of the Matsubara correlations when the coupling to the environment is very weak.

On the other hand, when the coupling with the environment starts to be a significant (but still perturbative) fraction of the system eigenfrequency, hybridization effects between the system and the bath become relevant.  As it will be shown in a later section, {\em the Matsubara correlations are essential to be able to correctly model both  the non-Markovian and the equilibrium properties in this parameter regime} (and which, in the weak-coupling case, were encoded in the detailed balance condition).  We first describe the HEOM, and how the Matsubara term can be included, even at zero temperature, with a fitting approach.

\section{The hierarchical equations of motion}

The HEOM method can in principle describe the exact behavior of the system in contact with a bosonic environment, without approximations. The derivation can be found in \cite{tanimura1989time, ishizaki2005quantum} and \cite{Ma_2012}, and the general procedure can be described as follows. Using the Gaussian properties of the free bath, one can write down a formally exact time-ordered integral for the reduced state of the system (or equivalently, a path-integral representation). This is difficult to solve directly. However, by assuming that the free bath correlation functions take can be written as a sum of exponentials, one can take repeated time derivatives to construct an exact series of coupled equations describing the physical density matrix, and auxiliary ones encoding the correlations between system and environment. These can be truncated at a level that gives convergent results.

The problem then lies in parameterizing the correlation functions of a given physical bath with a sum of exponentials. In practise one can either fit  \cite{duan2017zero, dattani2012optimal, fruchtman2016perturbative}  the correlation functions directly with exponentials, or fit the spectral density using a sum of overdamped (Drude-Lorentz) or underdamped Brownian motion spectral densities  \cite{tannor-meier, Kreisbeck2012, Chen2015}.  However, for the latter, as one might expect from the discussion so far,  the Matsubara frequencies in \eqref{matsubaracorr}  become increasingly important at low-temperatures \cite{strumpfer2012open}. These frequencies, in the HEOM, are numerically challenging to take into account due to the increasing number of auxiliary density operators \cite{moix2013hybrid, tang2015extended} (though using an alternative Pad\'{e} decomposition with the HEOM has been explored as a way to optimally capture the influence of these terms \cite{optimHEOM}).

In the zero-temperature ($\beta\rightarrow\infty$) limit, the Matsubara frequencies $\omega_k=2\pi k/\beta$ approach a continuum, i.e., $2\pi/\beta \to dx\to 0$ for  ${2 \pi k}/{\beta}\rightarrow x$. As a consequence, we can represent the Matsubara correlation in Eq.~(\ref{matsubaracorr}) as the integral

\begin{equation}
\label{eq:infinite_sum}
M(t) = - \frac{\gamma \lambda^2}{\pi} \int_{0}^{\infty}dx\frac{x e^{-xt}}{\left[ (\Omega+i\Gamma)^2 +x^2\right] \left[ (\Omega-i\Gamma)^2  +x^2 \right]}
\end{equation}

\begin{figure}[h]
\centering
\includegraphics[width = \columnwidth]{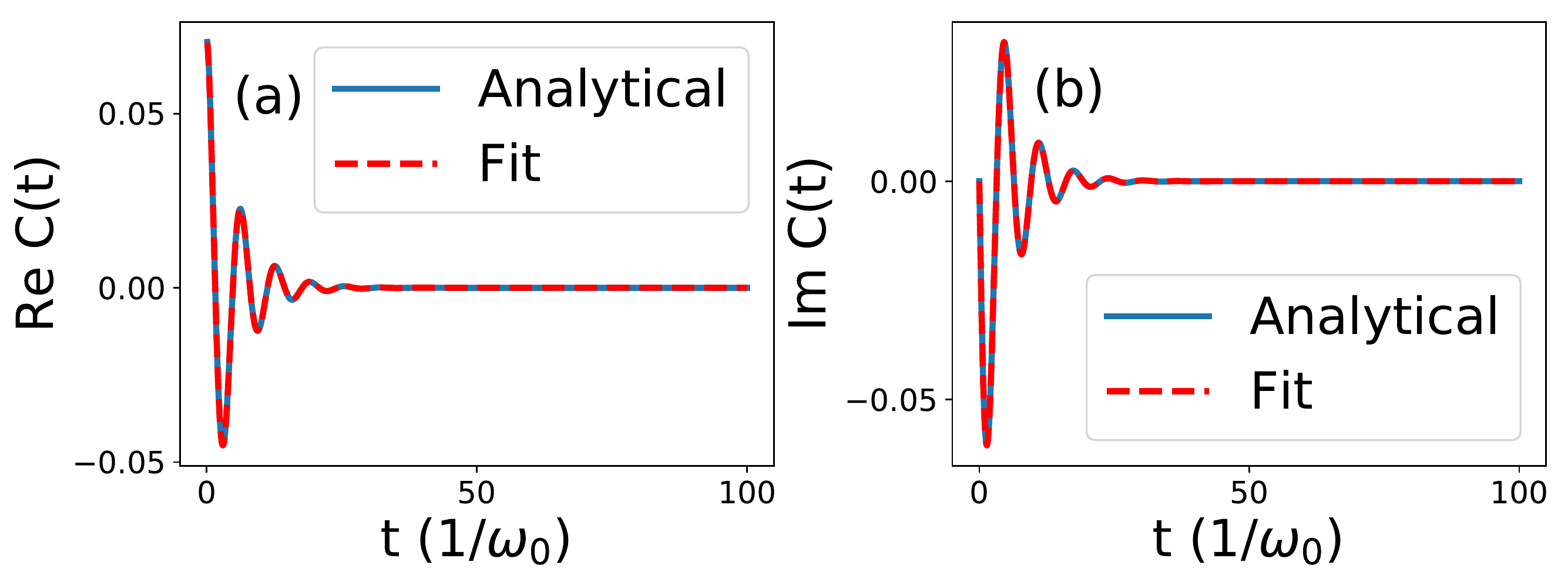}
\includegraphics[width = \columnwidth]{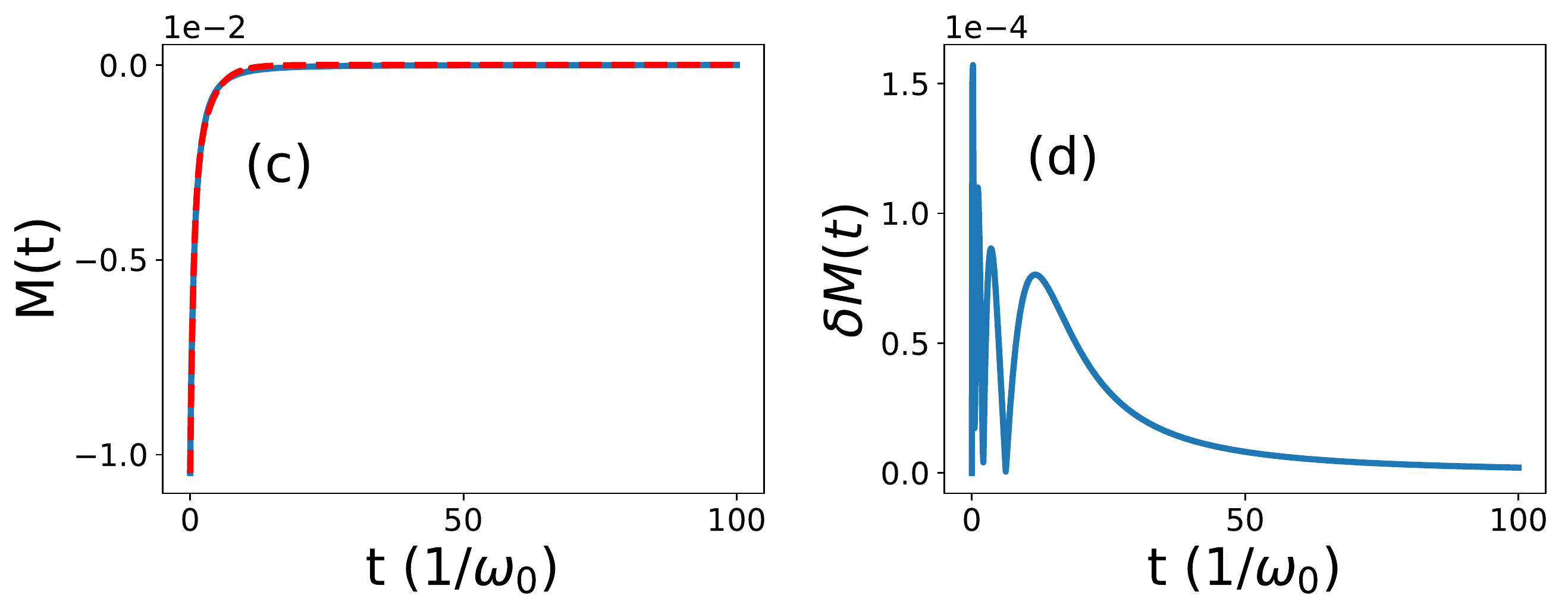}
\caption{The top two panels show (a) real and (b) imaginary parts of the correlation function for the underdamped Brownian motion spectral density with $\lambda = 0.4\omega_0, \gamma = 0.4\omega_0$, $T=0$.  The blue solid curves show the analytical formula from  \eqref{corrspecial} and the red dashed curves show the reconstruction of the same using four exponentials. Two of the exponents are given by the Matsubara fitting and the other two by the analytical non-Matsubara formula \eqref{eq:nonmatsubara}. In the bottom left panel (c) we explicitly plot the Matsubara part of the correlation function $M(t)$ alone, and its fit \eqref{biexp}. 
 The error in the fit is shown in the bottom right panel (d), which is also the same as the error in the real part of the correlation function. The imaginary part is exact and has no error after the reconstruction.}
\label{fig:fitting}
\end{figure}

However, this integral representation does not give a direct solution in exponential form. Using a fitting procedure we have found that we can capture the influence of these terms with the minimum of a biexponential function,
\bc
\beq
M_{\text{biexp}}(t) = c_1 e^{-\mu_1 t} +c_2 e^{-\mu_2 t}
\label{biexp}
\eeq
\ec
where $c_m$ and $\mu_i$ are real (for the choice of Matsubara decomposition we use here). Adding more terms increases the accuracy of the fit only marginally. In addition, each exponent leads to an added factorial level of complexity in using the HEOM method, thus one would like to keep the number of exponents to a minimum. In Figure \ref{fig:fitting}, we give an example of the fitting of the correlation function.

Given the above decomposition, we can finally write the full equations of motion.   However, the non-Matsubara terms given in \eqref{eq:nonmatsubara}  produce four exponents when decomposed into real and imaginary parts and employed with a fully generic formulation of the HEOM \cite{fruchtman2016perturbative}. It is more numerically convenient to reduce these to two exponents, following \cite{Ma_2012}, by defining (for $\beta=\infty$ for notational simplicity) the new parameters $c_3 = \frac{\lambda^2}{4\Omega}\left(1 - i\right)$, $c_4 = \frac{\lambda^2}{4\Omega}\left(1 + i\right)$, $\mu_3 = -i\Omega + \Gamma$, and $\mu_4 = i\Omega + \Gamma$.
Meanwhile, as described above, the Matsubara terms are entirely real, and given by \eqref{biexp}.

In the HEOM itself we denote the physical and auxiliary density matrices as $\rho_{\bar n}$ where $\bar n = [n_1, n_2, ..,n_K]$, (where here $K=4$), is a multi-index composed of non-negative integers $n_k$. The physical density matrix of the system, traced over the environment, is given by $ \rho_{\bar 0} = \rho_{[0, 0, ..., 0]} \equiv \mathrm{Tr}_{E}(\rho_{T})$.  Any other index denotes an auxiliary density operator  which encodes the correlations between system and environment, as we will discuss later. 
 We use $\rho_{\bar {n}_{k^{\pm}}} $ to denote a higher order ADO which differs from $\rho_{\bar n}$ in the $k^{th}$ index by $\pm 1$. For instance, $\rho_{0_{2^{+}}} = \rho_{[0, 1, 0, ..., 0]}$. The equations of motion given by HEOM  can be compactly written in the Liuoville space as the,

\beq
\dot \rho_{\bar n} = (-i\mathcal{L} - \sum_{k=1}^K n_k \mu_k )\rho_{\bar n} - i \sum_{k=1}^K \left (\mathcal{L}^{-}_k \rho_{\bar n_{k^{-}}} + \mathcal{L}^{+}_k \rho_{\bar n_{k^{+}}} \right) \nonumber \\
\label{eq:heom-eq}
\eeq
where $\mathcal{L} \rho = [H_s, \rho]$ and the $\mathcal{L}^{\pm}_k$ are Liouvilie space operators depending on the spin-bath coupling operator and the exponential decomposition of the correlation function \cite{tanimura1989time, ishizaki2005quantum}  given by $\mathcal{L}_k^{-} \rho_{\bar{n}_{k^{-}}} = n_k(c^R_k [Q, \rho_{\bar n_{k^-}}] + c^I_k \{Q, \rho_{n_{k^{-}}} \})$ and $\mathcal{L}_k^{+} \rho_{\bar{n}_{k^{+}}} = [Q, \rho_{n_{k^+}}]$.
Note again that this is not a generic construction \cite{fruchtman2016perturbative}, but is specific for the choice of decomposition of correlation functions we use here.

\subsection{Environment as a discrete set of modes}

Before discussing results predicted by the HEOM, it is useful to consider two complementary methods based on discrete decompositions of the environment. The idea that the behavior of an infinite continuum environment can be described by a finite set of discrete modes arises in both the methodology of ``pseudo-modes" \cite{Garroway,immoPM,Hughes:18,hughesPRL,plenio2018} and the so-called ``reaction coordinate mapping" \cite{Garg_1985, Martinazzo_2011,iles2014environmental}.  The former is based on the identification of frequencies in the correlation functions which are then assigned to a set of ``unphysical pseudo-modes" \cite{Garroway,plenio2018}.

In contrast, the reaction-coordinate (RC) method is instead based on a formal mapping of the full Hamiltonian environment Hamiltonian to a single ``reaction coordinate" and a residual (perturbative) environment.

\subsubsection{Pseudomodes model}
 As shown in the seminal work of Garraway \cite{Garroway} (and recently confirmed and generalized by Tamascelli {\em et al.} \cite{plenio2018,Lemmer_2018}), as long as the free correlation function of a discrete set of modes accurately reproduces  the correlation function of the full bath, their effect on a given system should be identical, a concept that recalls in spirit Baudrillard: {\em ``The simulacrum is never that which conceals the truth--it is the truth which conceals that there is none. The simulacrum is true.''}\cite{simul}.

 From the discussion so far, and the generalized proof in \cite{plenio2018}, it is evident that we can capture the full correlation function of the free environment \eqref{corrspecial} with a single under-damped mode for the non-Matsubara part \eqref{eq:nonmatsubara}, and two additional modes, from the fitting procedure \eqref{biexp}, which capture the Matsubara frequency contributions \eqref{matsubaracorr}. By construction, at zero temperature, the resulting dynamics of the system coupled to these effective modes should obey the total Hamiltonian,
 \beq
\label{eq:pmhamiltonian}
H_{\text{pm}} &=& \frac{\omega_q}{2}\sigma_z + \frac{\Delta }{2} \sigma_x + \sigma_z \sum_{i=1}^3\lambda_i (  a_i +   a_i^{\dagger}) + \sum_{i=1}^3\zeta_i a_i^{\dagger}a_i \nonumber \\
\eeq
Here, $\zeta_1=\Omega$, $\Omega=\sqrt{\omega_0^2 -\Gamma^2}$, $\zeta_2 = \zeta_3 = 0$, $\lambda_1 = \lambda/\sqrt{2\Omega}$, $\lambda_2 = \sqrt{c_1}$, $\lambda_3 = \sqrt{c_2}$  (where $c_1$ and $c_2$ are the  coefficients of the fitted Matsubara terms in \eqref{biexp}, and $\zeta_2 = \zeta_3 = 0$ because \eqref{biexp} contains no oscillating components).
 \par
\bc The damping of each pseudo-mode \ec is simply described by a Lindbladian with the corresponding loss rate,
\beq
\label{eq:lindblad}
D_i[a_i] =\mathcal{G}_i (2a_i\rho a_i^{\dagger} - a_i^{\dagger}a_i\rho - \rho a_i^{\dagger}a_i),
\eeq
where $\mathcal{G}_1= \Gamma$, $\mathcal{G}_2 = \mu_1$, $\mathcal{G}_i=\mu_2$.\\

Note that the couplings $\lambda_2$ and $\lambda_3$ between the pseudomodes associated with the Matsubara terms and the system are complex (since $c_1$ and $c_2$ are required to be negative), and thus the above Hamiltonian {\em is strangely non-Hermitian}.  This situation is not immediately covered by the general proof in \cite{plenio2018}. 
We extend their proof in \appref{APP:PM}, and show that, to properly take into account the negative $c_1$ and $c_2$, {\em the dynamics of the system has to be computed by solving the following \emph{pseudo}-Schrodinger equation } for the density matrix $\rho$
\begin{equation}\label{pse}
\frac{d}{dt}\rho=-i[H_\text{pm},\rho]+D[\rho]\;\;.
\end{equation}
where $D[\rho]=\sum_{i=1}^3 D_i[a_i]$. The adjective \emph{pseudo} not only refers to the pseudomodes in question, but also to the fact that, when  $H_\text{pm}$ is non-Hermitian, we are purposely \emph{not} taking the Hermitian conjugate when $H_\text{pm}$ acts on the right of $\rho$.

While we refer to \appref{APP:PM} for a detailed justification, given the non-Hermitian nature of the Hamiltonian in Eq.~(\ref{eq:pmhamiltonian}), it is worth presenting here a sketch of the proof.

Following a parallel strategy to the one presented in \cite{plenio2018}, it is possible to show that the dynamics of observables in the system+pseudomodes space (obtained by solving the pseudo-\emph{Lindblad} equation above),  is equivalent to a reduced pseudo-\emph{unitary} dynamics in which each pseudomode is coupled to  a bosonic environment under a rotating wave approximation and with a constant spectral density (defined for both positive and negative frequencies). 

As mentioned, the prefix pseudo- refers to the fact that the Hermitian conjugate is never taken when considering the Schr\"{o}dinger equation for the density matrix. From this auxiliary model, the reduced \emph{system}'s dynamics can be obtained through a Dyson equation. When the pseudomodes and their environments are in an initial Gaussian state, this equation is fully specified  by the two-time correlation function of the coupling operator $\sum_{i=1}^3\lambda_i (a_i+a_i^\dagger)$.

The advantage of considering an \emph{non-Hermitian} Hamiltonian together with a \emph{pseudo}-Schr\"{o}dinger equation in this derivation is that, by doing so, the Dyson equation for the reduced dynamics of the system is formally equivalent to one where the system is \emph{physically} interacting with a single environment via a \emph{Hermitian} coupling operator characterized by the same correlation function $C_0(t)+M_\text{biexp}(t)$. This completes the proof. 

To summarize, the reduced system dynamics computed from \eqref{pse} is equivalent to that of the original spin-boson model, \eqref{spinh}, under the assumption (or, in our case, approximation, due to the fitting procedure used to capture the Matsubara terms) that the correlation in Eq.~(\ref{corrspecial}) has the form,
\begin{equation}
C(t)=C_0(t)+M_\text{biexp}(t)\;\;.
\end{equation}

Remarkably, we will see in a later section that {\em this three-mode model precisely reproduces the results of the HEOM model, both when the Matsubara frequencies (modes) are neglected, and when they are included, and that they also allow for an interpretation of the auxiliary density matrices in the HEOM.}  In addition, the latter suggests that {\em the HEOM can be derived}, in some cases, {\em from the pseudo-mode model itself}  (akin to the ``dissipaton'' model introduced by Yan \cite{deom}). \par

We finish this section with a brief note on the effect of neglecting the Matsubara correlations, i.e., in considering the approximation $C(t)\mapsto C_0(t)$. In this case, only a single pseudomode is needed, i.e., $i=1$ in Eq.~(\ref{eq:pmhamiltonian}) and Eq.~(\ref{eq:lindblad}). Alternatively, as we show in the Appendix, this single pseudomode can be understood as mediating the interaction between the system and a residual bath of bosonic modes (with annihilation operator $f_k$ and frequency $\omega'_k$) with the Hamiltonian
\begin{equation}
\label{eq:rchamiltonian_pm}
\begin{array}{lll}
H_{\text{Mats}} &=& \displaystyle\frac{\omega_q}{2}\sigma_z + \frac{\Delta }{2}  \sigma_x + \lambda\sigma_z \frac{(a_1 + a_1^{\dagger}) }{\sqrt{ 2 \Omega }} + \Omega  a_1^{\dagger}a_1 \\
&+& \displaystyle\sum_k \omega'_k f_k^{\dagger} f_k +\displaystyle\sum_k  \frac{g'_k}{\sqrt{2  \Omega  } \sqrt{2\omega'_k }}  \left(f^\dagger_k a_1+a_1^\dagger f_k\right).
\end{array}
\end{equation}
where the couplings $g'_\alpha$ describing the interaction with the residual environment are characterized by the spectral density $J_\text{Mats}(\omega)=\gamma \Omega $ and defined for both positive and negative frequencies. This system has an interesting relation to another technique used to model the spin boson model: the reaction coordinate mapping.

\subsubsection{Reaction coordinate (RC) mapping}
 Returning to the full spin-boson Hamiltonian, in the reaction coordinate approach a unitary transformation maps the environment to a single-mode ``reaction coordinate'' and a residual bath. As discussed in \cite{iles2014environmental,Iles_Smith_2015,Strasberg_2016}, for the underdamped Brownian motion spectral density the new Hamiltonian is

\beq
\label{eq:rchamiltonian}
H_{\text{RC}} &=& \frac{\omega_q}{2}\sigma_z + \frac{\Delta }{2}  \sigma_x + \lambda\sigma_z \frac{(a+a^\dagger)}{\sqrt{ 2 \omega_0 }} + \omega_0 a^{\dagger}a \nonumber \\
&+& \sum_k \omega''_k d_k^{\dagger} d_k + \left(a+a^{\dagger}\right)\sum_k \frac{g''_k \left(d_k+d_k^{\dagger}\right)}{\sqrt{2  \omega_0  } \sqrt{2\omega''_k
 }}  .
\eeq
where the residual bath, described by operators $d_k$, with frequencies $\omega''_k$ and couplings $g''_k$,  has an Ohmic spectral density $J_{\mathrm{res}}(\omega) = \gamma \omega$. Given this new frame, for small $\gamma$ such that a Born-Markov-secular approximation for the residual bath is valid, one can derive a new master equation which describes the dynamics of the system coupled to the reaction coordinate, and which preserves detailed balance by definition [see \eqref{eq:rcbms_2} in \appref{RCME}].

Conversely, it is interesting to understand what set of approximations in the RC model are equivalent to neglecting the Matsubara correlations as in Eq.~(\ref{eq:rchamiltonian_pm}). To achieve this goal, we adapt the intuitive procedure outlined in \cite{ingold2002path} (see also \cite{Lemmer_2018}).  To start, we rewrite the spectral density in Eq.~(\ref{underdamp}) as a sum of two Lorentzians
\begin{equation}
\label{eq:spectralDensityIngold}
J(\omega) =  \frac{ \gamma \lambda^2}{4 \Omega}  \left[\frac{1}{(\omega-\Omega)^2+\Gamma^2} -\frac{1}{(\omega+\Omega)^2+\Gamma^2} \right].
\end{equation}
We now consider the effects of rotating-wave and Markov approximations in computing the correlations  in Eq.~(\ref{corrspecial}).

Intuitively, the rotating wave-approximation neglects terms in which the system decays to a lower state by absorbing energy from the bath (or vice versa) while the Markov approximation (for the interaction between the RC and the residual bath)  replaces weak frequency dependencies with their value at resonance. Furthermore, we need to consider that, from the analysis of Eq.~(\ref{eq:rchamiltonian_pm}), the residual bath should have both positive and negative frequencies.

In order to impose the rotating-wave approximation (\cite{ingold2002path,Loudon}) at positive (negative) frequencies, we neglect the peak at negative (positive) frequencies in the spectral density, i.e., the second (first) term in Eq.~(\ref{eq:spectralDensityIngold}). With this in mind, by inserting Eq.~(\ref{eq:spectralDensityIngold}) into Eq.~(\ref{corrspecial}),  we obtain
\begin{equation}
\begin{array}{lll}
C(t) &\simeq& \displaystyle\frac{\lambda^2 \gamma}{8 \pi \Omega}\int_{-\infty}^\infty d\omega \frac{\coth[\beta \Omega/2]\cos{\omega t - i\sin{\omega t}}}{(\omega-\Omega)^2+\Gamma^2} \\
&&-\displaystyle\frac{\lambda^2 \gamma}{8 \pi \Omega}\int_{-\infty}^\infty d\omega \frac{\coth[-\beta \Omega/2]\cos{\omega t - i\sin{\omega t}}}{(\omega+\Omega)^2+\Gamma^2} \\
&=& \displaystyle \frac{\lambda^2}{2\Omega}  e^{-\Gamma t}e^{-i\Omega t}\\
\end{array}
\end{equation}
where, in the first step, we both approximated the value of the hyperbolic cotangent at the resonant values $\pm\Omega$, enforcing the Markov approximation \cite{ingold2002path}, and set $\beta\rightarrow \infty$.

This correlation function is the same as the non-Matsubara part in Eq.~(\ref{eq:nonmatsubara}) for $\beta \to \infty$.
Thus, in the context of the RC Hamiltonian, when we ignore the Matsubara terms, we are performing both a rotating-wave approximation and Markov approximation on the interaction between the {\em collective mode} and the {\em residual environment}. Note that these considerations, while shedding  intuition upon the relation between the two models in  Eq.~(\ref{eq:rchamiltonian}) and Eq.~(\ref{eq:rchamiltonian_pm}), should not be considered as a rigorous mapping (for example, the RC and pseudomode have different frequencies).

\begin{figure}[ht]
\includegraphics[width = \columnwidth]{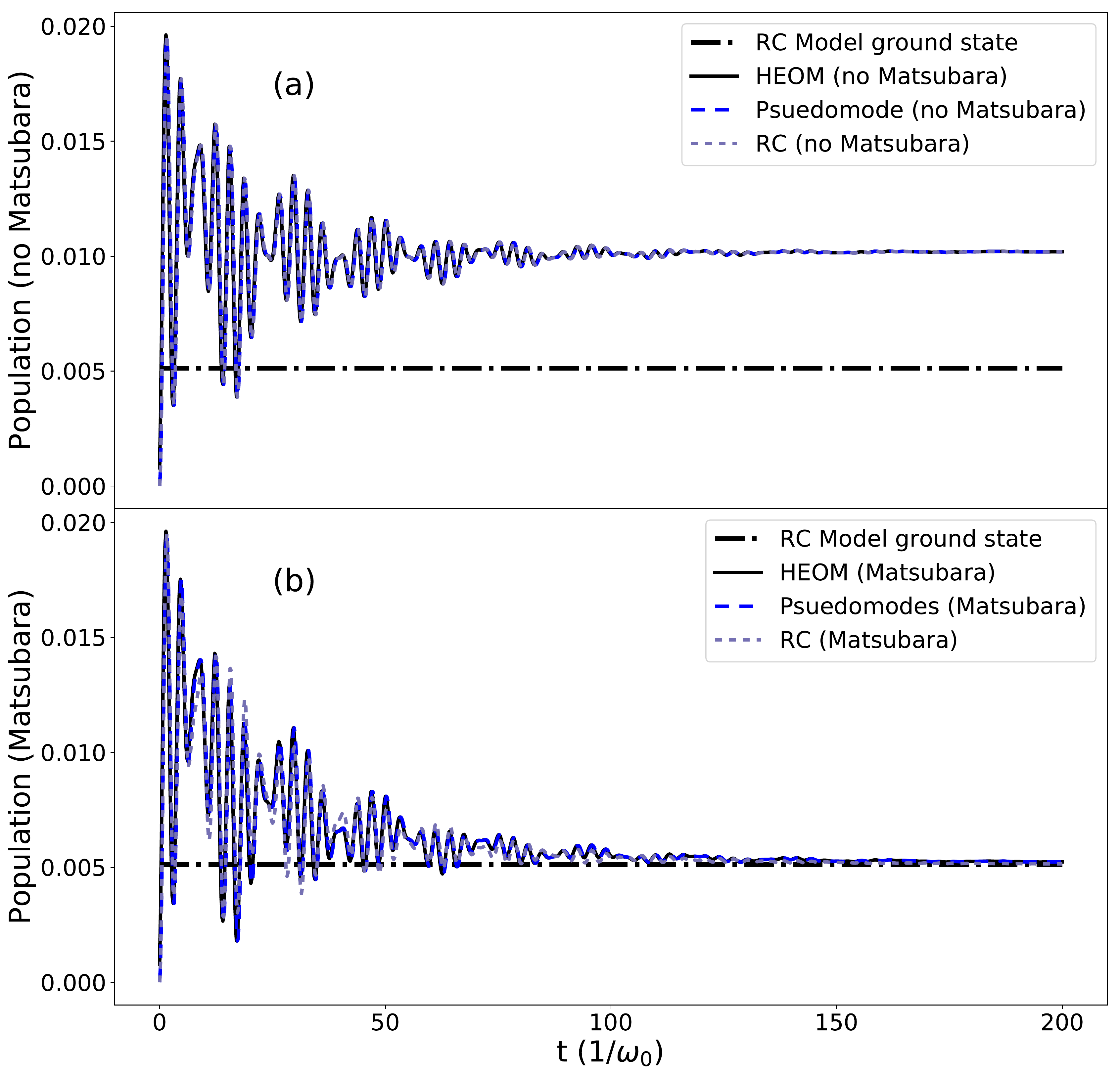}
\caption{Bath mode occupation for the various methods.  For the RC method this is directly the RC mode occupation $\expec{a^{\dagger}a}$. For the HEOM and the pseudomode methods this is the occupation of the effective mode associated with the frequency $\Omega$.  The parameters are  $\lambda = 0.2\omega_0$, $\gamma = 0.05\omega_0$, $\omega_q = 0$, $\Delta=\omega_0$, $T=0$. The upper panel (a) gives the results of the three models we consider without Matsubara terms (both direct, and effective in the RC case). For this choice of parameters all three models coincide. In the lower panel (b) we show the three models with Matsubara terms included, and all three tend towards to a steady-state which corresponds to the ground state of $H_{RC}$ (dashed-dotted black line).}\label{4}
\end{figure}

\begin{figure}[ht]

\includegraphics[width = \columnwidth]{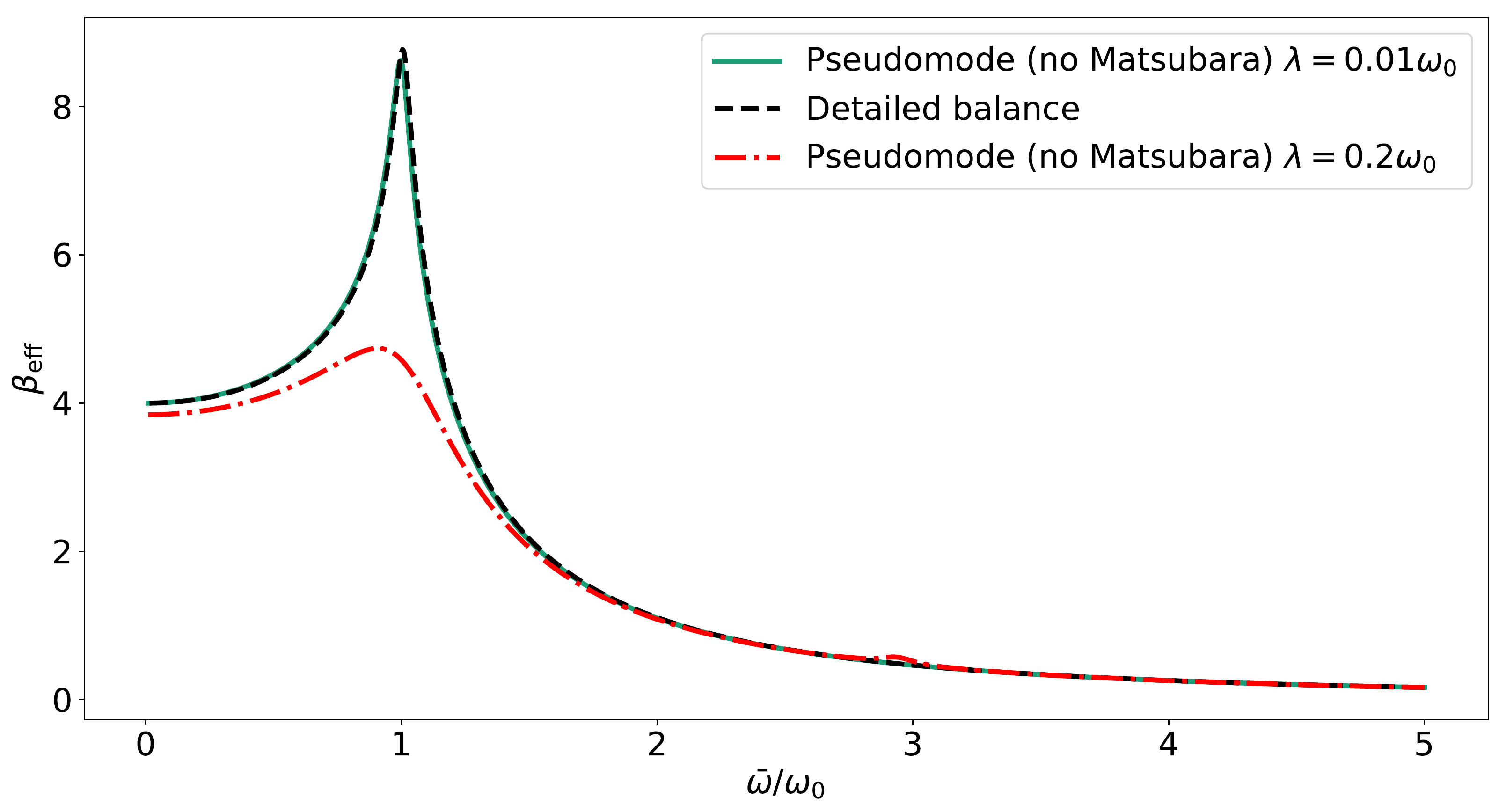}
\caption{The effective inverse temperature $\beta_{\mathrm{eff}}$ extracted from the steady-state populations of the qubit, using the pseudo-mode method without Matsubara corrections. For weak coupling, $\lambda = 0.01\omega_0$, we see that the effective inverse temperature fits closely that given by the detailed balance consideration in \eqref{beffeq}, and arises due to the neglect of the Matsubara terms.  As the coupling is increased to $\lambda = 0.2\omega_0$, we see that the effective inverse temperature decreases relative to that predicted by \eqref{beffeq}, due to hybridization between system and environment. }\label{4b}
\end{figure}

\begin{figure}[ht]
\includegraphics[width = \columnwidth]{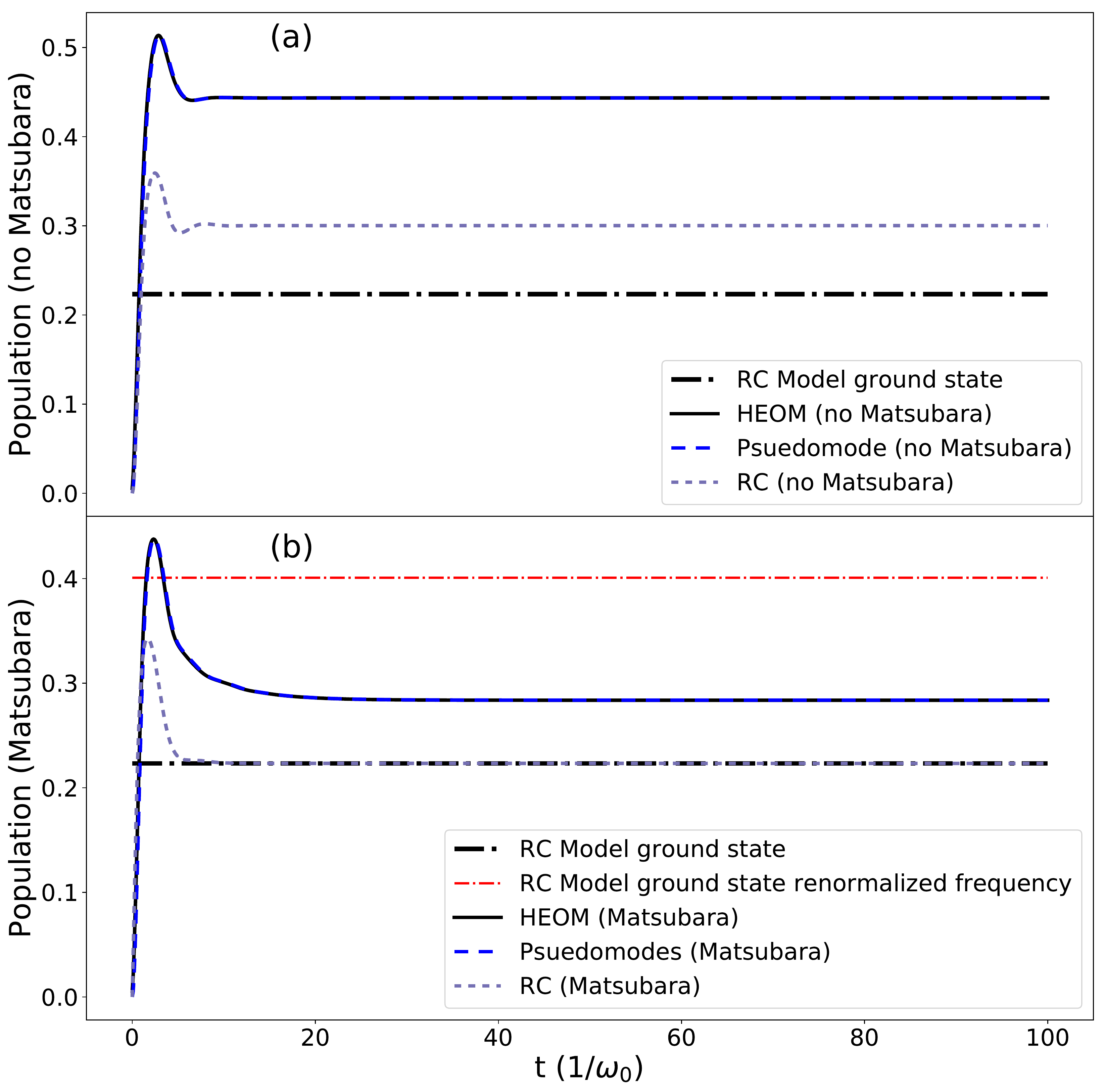}
\caption{\label{fig:bath-occupation}Bath mode occupation for a strongly coupled broad bath $\lambda = \omega_0$, $\gamma = \omega_0$, and again set $\omega_q = 0$, $\Delta=\omega_0$, $T=0$. The upper panel (a) gives the results of the three models we consider without Matsubara terms (both direct, and effective in the RC case). For this choice of parameters all the HEOM and pseudomode models coincide, but the RC model shows some deviations it does not take into account the renormalized frequency $\Omega$. In the lower panel (b) we show the three models with Matsubara terms included, and now only the RC model tends to the ground-state of $H_{RC}$, while the pseudo-mode and HEOM models coincide and take into account corrections due to strong correlations with the effective `Matsubara modes' (note that the RC model is not corrected by just including the renormalized frequency, as shown by the red dot-dashed line, which shows the ground-state occupation for an RC model with a phenomenologically altered frequency, i.e., by setting the frequency of the RC mode in \eqref{eq:HRCAPP} to be equal to $\Omega$).}\label{5}
\end{figure}

Overall this suggests that the Matsubara frequencies play two roles: first of all, they restore detailed balance, both on the level of the system, in the weak-coupling regime (as expected), and also on the level of the system and RC mode, in the strong-coupling and narrow-bath regime.
 Secondly, beyond the weak-coupling and narrow-bath regime, they describe the non-negligible influence of `background' modes in the environment not captured by the reaction coordinate itself (e.g., strong correlations with the residual bath).

\section{Virtual excitations in the ground state}

An interesting phenomenon that arises in the ultrastrong coupling regime where the qubit-environment coupling $g_k$ is comparable to the bath frequencies $\omega_k$ in \eqref{spinh} is the appearance of virtual photons \cite{DeLIberato17}. In this scenario, the hybridized system-environment ``ground-state'' (which in principle should be the steady-state at zero temperature) contains a finite population of photons which cannot be directly observed (or emitted into ``other modes'' or environments).

In our treatment of the ultra-strong coupling regime, the Matsubara terms are crucial to get the correct photon population in a single collective mode, and trap that population. In order to show this, we first consider the RC picture where the collective bath coordinates are approximated with a single mode,
\beq
\sum_k \frac{g_k}{\sqrt{2 \omega_k}} (b_k^{\dagger} + b_k) \equiv \frac{g}{\sqrt{2 \omega_0}} (a^{\dagger} + a)
\eeq
This mapping gives a very clear picture of the dominant influence of the environment in terms of the collective RC mode, such that any virtual or real photon population {\em of the collective mode} is given by the expectation of the number operator, $\expec{a^{\dagger}a}$ (though this does not directly correspond to the original bath-mode occupation).  

Can a similar quantity be extract from the HEOM? It has been shown \cite{shiPosition, ShiMomentum}  that higher-order moments of the total bath coupling operator can be extracted from certain combinations of auxiliary density operators returned by the HEOM. Similarly, for a single undamped mode, \cite{schinabeck} showed that the population is given by the second level auxiliary density matrix.  In our case,  we can extract populations that correspond to precisely those of the pseudo-modes (see \appref{ADO}). For example, the occupation of the first pseudo-mode is given by
\beq\label{heompop}
\expec{a_1^{\dagger} a_1} = \rho_{1, 1, 0, 0}/\lambda_1^2.
\eeq
It is clear then that the ADOs and the pseudo-modes bear a close relationship.

As we can see in \figref{4}, (starting from the initial condition of a zero-temperature environment, and the qubit in the ground-state of the free system Hamiltonian), in the absence of the Matsubara terms, the population of the excited state of the two-level system (see \figref{panel} in the appendix), and the population of the $a_1$ mode predicted by the HEOM from \eqref{heompop} matches closely that of the RC model with the approximation of the RWA for the RC-residual bath coupling and a flat-residual-bath approximation (described by Eq.~(\ref{eq:rcingold})). In this case the population increases to a steady state which can be ascribed to the
artificial non-equilibrium situation induced by neglecting the Matsubara correlation (see \figref{4b} for a comparison of the resulting effective inverse temperature to that predicted by \eqref{beffeq}). In the RC model, without Matsubara contributions, because the state $\rho(t)$, of the qubit and RC mode, evolves through the Lindblad equation shown in Eq.~(\ref{eq:rcingold}), characterized by the bare annihilation operator $a$, the rate of energy dissipation into the residual environment is given by
\begin{equation}
J(t)=\gamma\omega_0~ \text{Tr}\left(a^\dagger a \rho(t)\right)
\label{emission1}
\end{equation}
i.e., proportional to the average photons in the steady-state. {\em However we know that this emission is unphysical, as it both violates detailed balance and energy conservation.}

In contrast, the addition of the Matsubara terms to the HEOM, the addition of the Matsubara modes to the pseudo-mode model, and the corresponding removal of the unphysical assumptions in the RC model, results in dynamics in all three cases which tend towards a steady state which is close to the ground state of the coupled system-RC Hamiltonian.  In this case the HEOM and pseudo-mode model match exactly, while the RC model gives qualitative agreement. This trend is one of our primary results: the addition of Matsubara terms to the HEOM (or equivalently Matsubara modes to the pseudo-mode model) restores detailed balance, and traps photons in an effective ground state, as confirmed by the RC model.  In this case the state $\rho(t)$ of the qubit and RC mode evolves through the Lindblad equation shown in Eq.~(\ref{eq:rcbms}) and Eq.~(\ref{eq:rcbms_2}) characterized by jump operators between eigenstates. As a consequence, since the steady-state is the ground-state, there is no steady-state energy dissipation (see \eqref{emission2}) into the residual bath.

As $\gamma$ is increased,  cf. \figref{5}, we see a  deviation between HEOM and RC models (see also \figref{fig:comparision-all} for a comparison of system populations).  For strong coupling and broad baths, the Matsubara terms become more relevant, as does the error arising from the fitting procedure.  In the appendix we perform an error analysis which suggests that the difference between the RC results and the HEOM results exceed potential errors arising from the fit. Thus, we primarily ascribe this difference to the breakdown of the perturbative approximation for the residual bath in the RC model, which becomes more pronounced as $\gamma$ is increased.

One might attribute the difference to the fact the RC model does not take into account the frequency shift that we see in \eqref{eq:nonmatsubara}.  However, phenomenologically solving for the ground state of the system coupled to an RC mode with renormalized frequency $\Omega$ actually predicts a larger population (shown by the red dot-dashed line in \figref{5}) than the normal system-RC ground-state due to the decreased frequency of the non-Matsubara mode \cite{DeLIberato17}.  The fact that this predicted population is also larger than the full HEOM/pseudo-mode results suggests that, as $\gamma$ is increased, the correlations between the system and the pseudo-modes associated with the Matsubara frequencies become stronger, and actually reduce the population in the non-Matsubara pseudo-mode \cite{DeLIberato17}.  However, without the RC model to guide us with a physical interpretation in this limit, it becomes difficult to associate the populations of the Matsubara modes to real physical modes, collective or otherwise \cite{Martinazzo_2011,chain1, chain2, chain3, Chin2013}.  In fact, as described earlier, since their contribution to the correlation functions of the bath is {\em negative} in the parameter regimes we consider here, in the pseudo-mode model their coupling to the system is {\em non-Hermitian}, accentuating their purely mathematical nature.

\section{Conclusion}
We have analyzed the dynamics and steady-state properties of the zero-temperature spin boson model in the strong and ultra-strong coupling regime using three different techniques. We showed that the Matsubara terms, taken into account with a fitting procedure in the HEOM and pseudo-mode methods, restore detailed balance, even in the ultra-strong coupling regime.  This was validated by a comparison to the reaction coordinate method, which also indicates the Matsubara terms are important for the correct `trapping' of virtual excitations in the collective ground-state.

Simultaneously, we showed that a pseudo-mode model can exactly capture the same dynamics as the HEOM, and can take into account negative contributions to the correlation functions, like the Matsubara frequencies, via a ``pseudo-Schr\"{o}dinger equation''. Our results also elucidate the relationships and differences between the three methods employed herein, particularly the strong relationship between pseudo-mode treatment and the HEOM.

Future work includes generalizing to arbitrary spectral densities for systems such as superconducting qubits coupled to transmission lines (with potentially structured environments \cite{anton}), and photosynthetic complexes \cite{Akihito09,panitchayangkoon2011direct, Peter11, Lambert_2013,Chen2015,ishizakireview,Scholes2017}. In addition, in the broad-bath limit, it may be possible to assign direct physical meaning to the ADOs of the HEOM, and the Matsubara modes of the pseudomode method, by comparison to bosonic-chain mappings of the environment \cite{Martinazzo_2011,chain1, chain2, chain3, Chin2013}, in the same way the RC mapping guides us in this work.  This might allow, for example, inspection of spatial dependencies of the photon population, as revealed by other methods \cite{Zueco2013, Munoz18}.  We hope that these new insights can help towards a better understanding of ultrastrong coupling at zero-temperature in continuum systems, and emphasize the impact of the positive frequency nature of many physical environments (and the resulting appearance of Matsubara frequencies).

\acknowledgements
We would like to thank Stephen Hughes for helpful suggestions on the pseudomode approach, and Ken Funo, David Zueco, Simone De Liberato, and Fabrizio Minganti for feedback and comments. S.A. was supported by the RIKEN IPA program. N.L. acknowledges support from JST PRESTO, Grant No. JPMJPR18GC. N.L. and F.N. acknowledge
support from the RIKEN-AIST Joint Research Fund and the Sir John Templeton Foundation. F.N. is partly
supported by the MURI Center for Dynamic Magneto-Optics
via the Air Force Office of Scientific Research
(AFOSR) (FA9550-14-1-0040), Army Research Office
(ARO) (Grant No. W911NF-18-1-0358), Asian Office of
Aerospace Research and Development (AOARD) (Grant
No. FA2386-18-1-4045), Japan Science and Technology
Agency (JST) (the Q-LEAP program, the ImPACT program
and CREST Grant No. JPMJCR1676), Japan
Society for the Promotion of Science (JSPS) (JSPSRFBR
Grant No. 17-52-50023, JSPS-FWO Grant No.
VS.059.18N).

\bibstyle{apsrev4-1}
\bibliography{references/references,references/rcrefs,references/more_refs}{}

\begin{appendix}
\section{Reaction coordinate (RC) mapping}\label{RCME}

The reaction coordinate (RC) mapping is described in detail in \cite{Garg_1985, Martinazzo_2011,iles2014environmental, Strasberg_2016}, and we will only discuss it briefly. After the mapping, one can derive an appropriate master equation description of the residual bath.  For the full RC-model to which we compare the HEOM results in the main paper we use a Born-Markov-secular master equation description of the residual bath (described by the $d_k$ modes in $\eqref{eq:rchamiltonian}$), which has the form:

\beq
\label{eq:rcbms}
\dot{\rho} = -i[H_{RC},\rho] +  D^{(1)}[\rho]
\eeq
where
\beq
\label{eq:HRCAPP}
H_{RC} &=& \frac{\omega_q}{2}\sigma_z + \frac{\Delta }{2} \sigma_x + \sigma_z \frac{\lambda}{\sqrt{ 2 \omega_0 }} (a + a^{\dagger}) + \omega_0 a^{\dagger}a \nonumber \\
\eeq
and
\beq
\label{eq:rcbms_2}
D^{(1)}[\rho] &=& \sum_{i,j>i}D_{i,j}[\rho]\\
D^{(1)}_{i,j}[\rho] &=& J_{\mathrm{res}}(\Delta_{i,j})\bar{X}_{i,j}\left[2\ket{\psi_i}\bra{\psi_j} \rho \ket{\psi_j}\bra{\psi_i} \right.\nonumber\\
&-& \left.\ket{\psi_j}\bra{\psi_j} \rho -\rho  \ket{\psi_j}\bra{\psi_j}\right].\nonumber
\eeq

Here $\Delta_{i,j}$ is the energy difference between the eigenstates $\psi_i$ and $\psi_j$ of $H_{RC}$.  In addition, $\bar{X}_{i,j} = |\bra{\psi_j}\hat{X}\ket{\psi_i}|^{2}$, $\hat{X} = (a+a^{\dagger})/\sqrt{2\omega_0}$.

This master equation is used to produce the purple dashed curves in Figure \ref{fig:bath-occupation}. For small values of $\gamma$ (narrow spectral densities) this qualitatively approximates the HEOM result. 
The master equation predicts an energy dissipation into the environment in the following form:
\begin{equation}
J(t)=\sum_{i,j>i} \Delta_{i,j} J_{\mathrm{res}}(\Delta_{i,j})\bar{X}_{i,j} \text{Tr}\left(\ket{\psi_j}\bra{\psi_j} \rho(t)\right).
\label{emission2}
\end{equation}

\subsection{RC with RWA and flat-bath spectral density}\label{RCME2}
We wish to see the effect of removing the Matsubara terms from the RC method. In the HEOM method, it is as straightforward as ignoring them in the correlation function. \bc However, \ec to produce an effective RC model which takes into account the neglect of Matsubara terms we have to turn to the series of approximations suggested by Ingold \cite{ingold2002path}. First,  the interaction in $\eqref{eq:rchamiltonian}$ between the RC mode and the residual bath is forced to obey a rotating=wave approximation (even though such an approximation is not justified).  Second, the residual bath spectral density is set as frequency independent such that $ J_{\mathrm{flat}}(\omega) = \gamma \Omega$. Applying both approximations, in addition to the standard Born-Markov-secular approximations, leads to the following master equation,
\beq
\label{eq:rcingold}
\dot{\rho} = -i[H_{RC},\rho] + D^{(2)}[\rho]
\eeq
where as before
\beq
H_{RC} &=& \frac{\omega_q}{2}\sigma_z + \frac{ \Delta}{2} \sigma_x + \sigma_z \frac{\lambda}{\sqrt{ 2 \omega_0 }} (a + a^{\dagger}) + \omega_0 a^{\dagger}a \nonumber \\
\eeq
and now
\beq
D^{(2)}[\rho] = \frac{\gamma}{2} [ 2 a\rho a^{\dagger} - a^{\dagger}a\rho -\rho a^{\dagger}a].
\eeq
Note  that in the two master equations in this section the frequency of the RC is found to be $\omega_0$.  However, the frequency of the primary oscillating-mode correlation function, and the corresponding pseudomode, is $\Omega=(\omega_0^2 - (\gamma/2)^2)^{1/2}$.  The difference arises because that renormalized frequency is exact to all orders, while the frequency for the RC mode master equation in contact with the residual bath is only approximate. \par

\begin{figure*}[ht]
\includegraphics[width = \columnwidth]{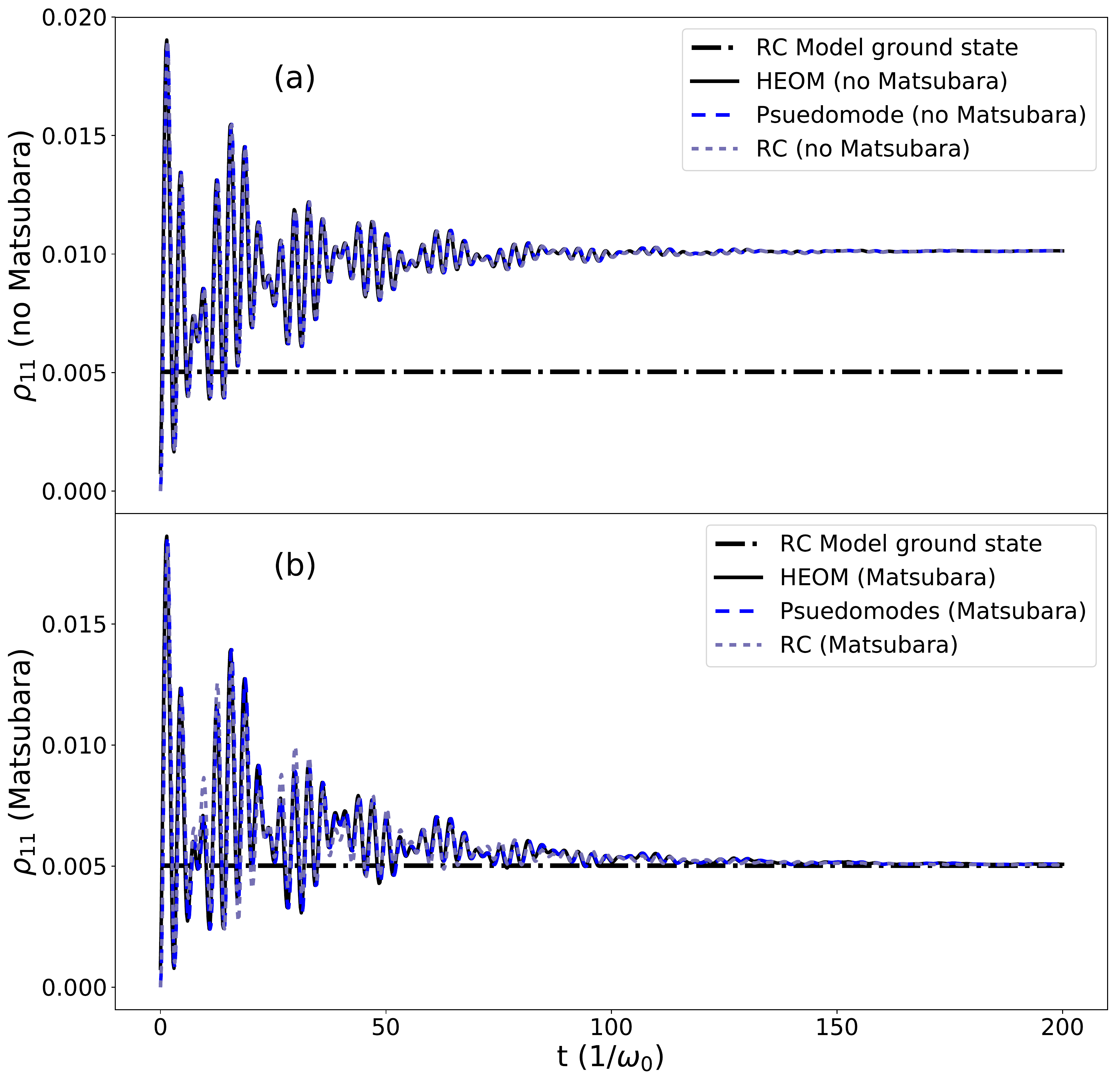}
\includegraphics[width = \columnwidth]{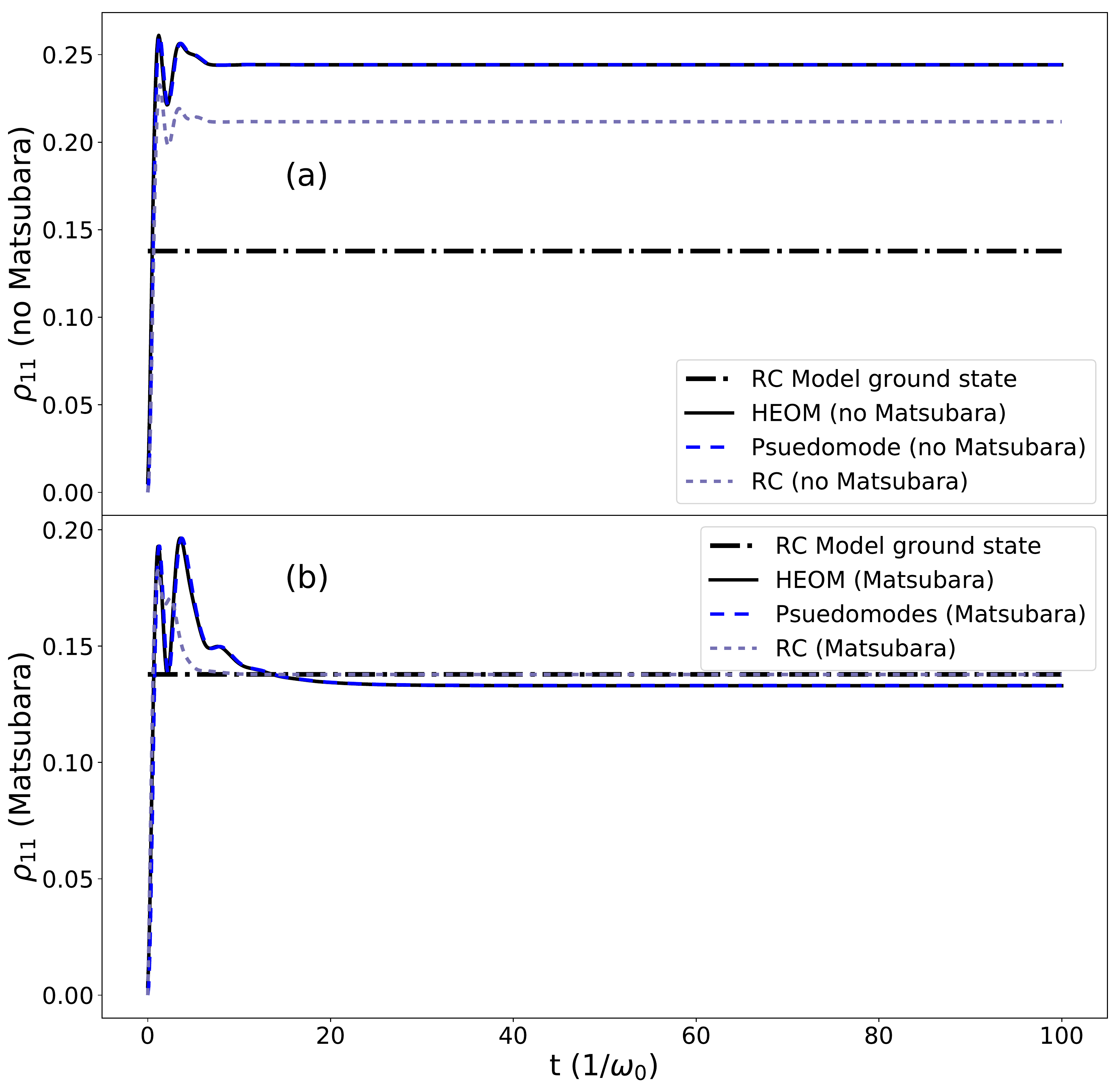}
\caption{
\label{fig:comparision-all}
Probability for the qubit to be in its excited state $\rho_{11}=\bra{1} \rho\ket{1}$, as given by different methods.  The left panels use the parameters $\lambda = 0.2 \omega_0$, $\gamma = 0.05\omega_0$, $\omega_q = 0$, $\Delta=\omega_0$, $T=0$, as in \figref{4}. 
The right panels use  $\lambda = \omega_0$, $\gamma = \omega_0$, as in \figref{5}.  The curves follow the same labelling scheme as \figref{4} and \figref{5}.}\label{panel}
\end{figure*}

 Now we can see that the results produced by this ``incorrect" derivation of the master equation are, for small $\gamma$, exactly the same as the one by the HEOM method where the Matsubara frequencies are ignored, see Figure (\ref{fig:comparision-all}).

\section{Virtual excitations from auxiliary density operators}\label{ADO}

Several works \cite{shiPosition, ShiMomentum} have explicitly shown how to extract moments of the bath coupling operator $X = \sum_k \frac{g_k}{\sqrt{2 \omega_k}} \left(b_k+b_k^{\dagger}\right)$ and the equivalent sum of mass weighted momenta, $P =i \sum_k g_k\sqrt{\frac{\omega_k}{2}}\left(b_k^{\dagger}-b_k\right)$ from the ADOs of the HEOM. In the limit of a single (and consequently undamped) mode in the environment, Schinabeck {\em et al} \cite{schinabeck} showed that the occupation of the (essentially single) bath mode can be extracted \bc from \ec certain second-level ADOs in the hierarchy.

In the general case, we can make progress by making a similar comparison between the HEOM and the equations of motion for the coupling operators \bc for \ec each pseudo-mode in \eqref{eq:pmhamiltonian}.  In the interaction picture, each mode operator rotates as $a_i(t) = a_i \exp{\left(- i\omega_i t\right)}$. The equation of motion for $\mathrm{Tr}_E[\lambda_i a_i(t)\rho(t)]$, derived from the Lindbladian master equation given in \eqref{eq:pmhamiltonian} and \eqref{eq:lindblad}, follows as,
\beq
&&\frac{d}{dt}\mathrm{Tr}_E[\lambda_i  a_i(t)\rho(t)]  \\
&=& (-i\mathcal{L} - i\Omega - \Gamma)\mathrm{Tr}_E[\lambda_i  a_i(t)\rho(t)] \nonumber\\
&-&i \left[\sigma_z \quad \mathrm{Tr}_E[\lambda_i  a_i(t)\left(\sum_k \lambda_k\{a_k(t)+ a_k(t)^{\dagger}\}\right)\rho(t)]\right.\nonumber\\
&-& \left. \mathrm{Tr}_E[\left(\sum_k \lambda_k\{a_k(t)+ a_k(t)^{\dagger}\}\right)\lambda_i a_i(t)\rho(t)]\sigma_z\right].\nonumber
\eeq
Here, $\mathcal{L}\rho = -i[H_s,\rho]$. \bc We can compare \ec this to the equation of motion of $\rho^{0,0,0,1}$ in the HEOM as per \eqref{eq:heom-eq},
\beq
\frac{d}{dt}\rho^{0,0,0,1}&=& (-i\mathcal{L} -i \Omega - \Gamma)\rho^{0,0,0,1} \\
&-&i \left[\sigma_z \sum_k P_+^k \rho^{0,0,0,1}-  P_+^k \rho^{0,0,0,1}\sigma_z\right].\nonumber
\eeq
where the operator $P_+^k\rho^{n}=\rho^{n_k+1}$  raises the $k^{th}$ element of $n$ by one.   Similar equations can be derived for the other first-tier \bc ADOs and we can \ec immediately make a correspondence between the two equations, such that $\expec{a_1^{\dagger} a_1} = \rho_{0, 0, 1, 1}/\lambda_1^2$.
Note that the non-Matsubara pseudo-mode is associated with the last two indices, corresponding to $a_1(t)$ and $a_1(t)^{\dagger}$, while the two Matsubara modes, being zero frequency modes,
are just associated with a single index each.

\section{Error bounds from fitting}
The error due to the numerical fitting of the infinite Matsubara sum with the biexponential in \eqref{biexp} will inevitably lead to an error in the dynamics of the system. This has been discussed extensively in Mascherpa et al. \cite{PlenioErrorsCorr} where it was argued that an error in the correlation function, $\Delta C(t)$, leads to a corresponding error in the expectation of any operator which  is bound by the inequality,
\begin{eqnarray}\label{plenioineq}
\mid\!\Delta \langle \hat {\mathcal{O}}(t) \rangle\!\mid \; \le \; \mid \mid \hat {\mathcal{O}} \mid \mid \left( e^{ \int_0^{t} dt^{\prime} \int_0^{t^{\prime}} dt^{\prime \prime} \mid \Delta C(t^{\prime} - t^{\prime \prime})\mid } - 1 \right)
\end{eqnarray}
where $\mid \mid \hat {\mathcal{O}} \mid \mid$ denotes the operator norm. In this section we consider whether this result is useful to characterize the error in the dynamics of our model in the main text.  Before showing that result, we first discuss another comparison we will make: the exactly solvable pure-dephasing model.

\begin{widetext}
\begin{figure*}[ht]
\includegraphics[width = 18em]{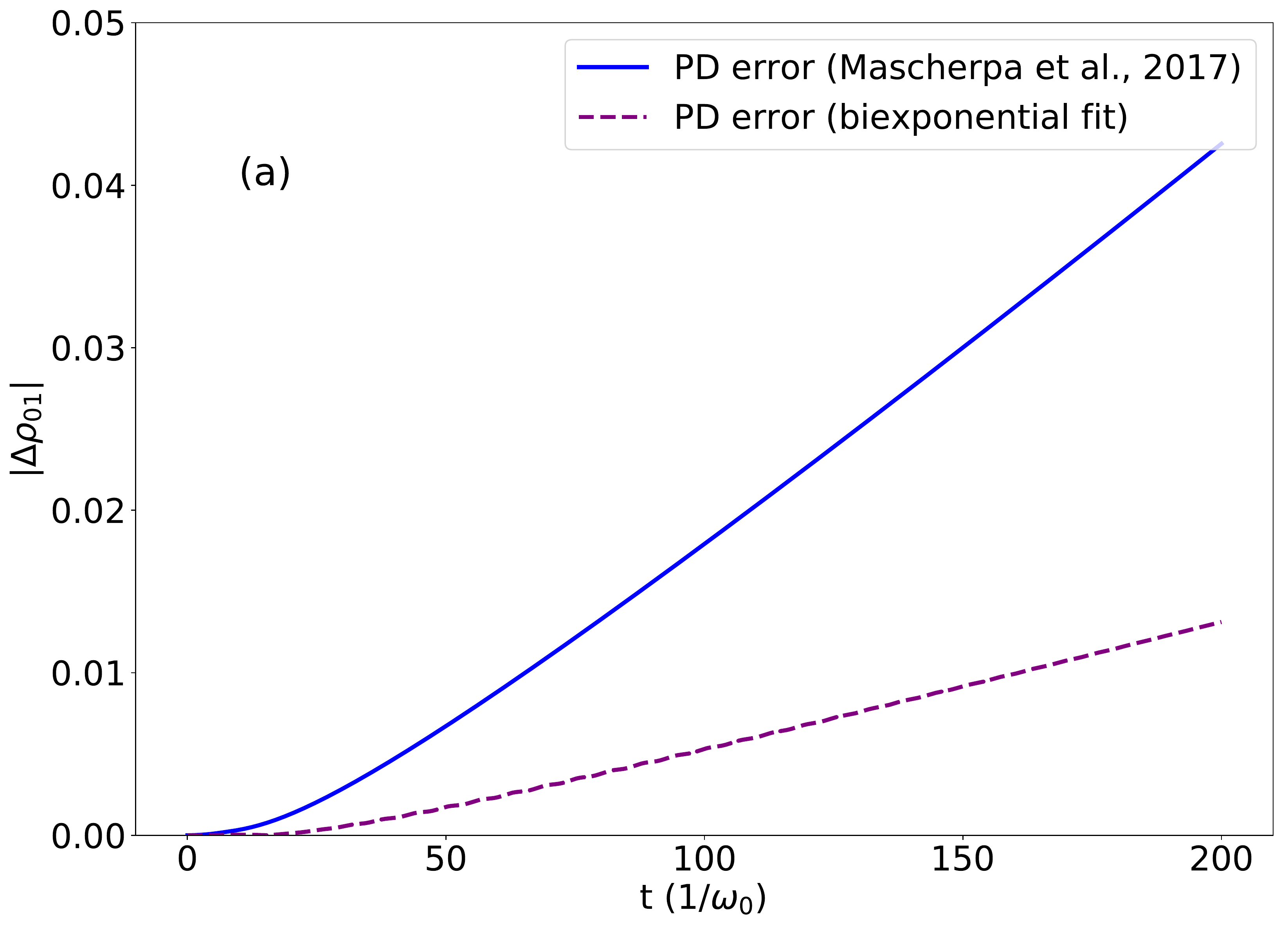}
\includegraphics[width = 18em]{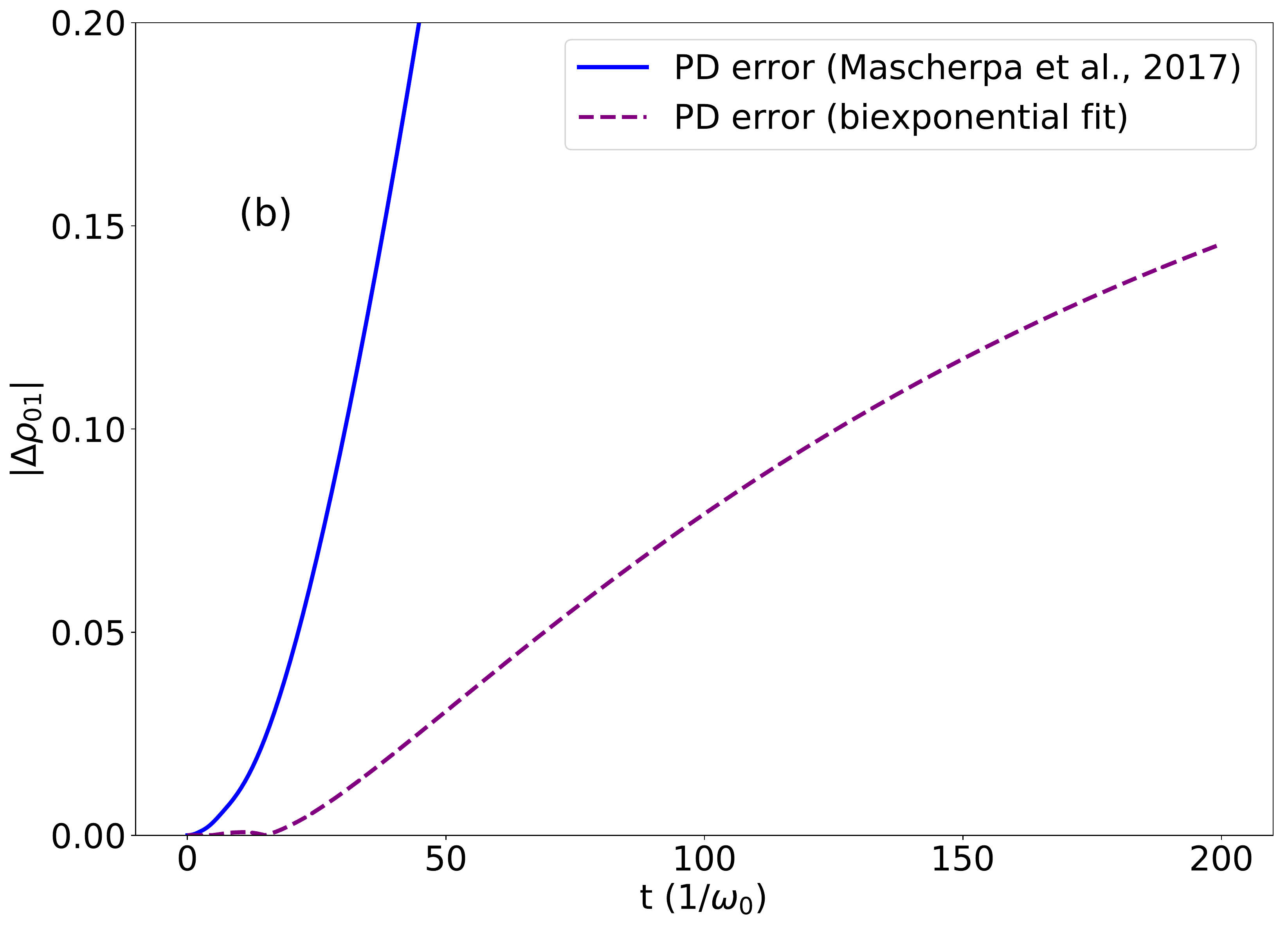}
\includegraphics[width = 18em]{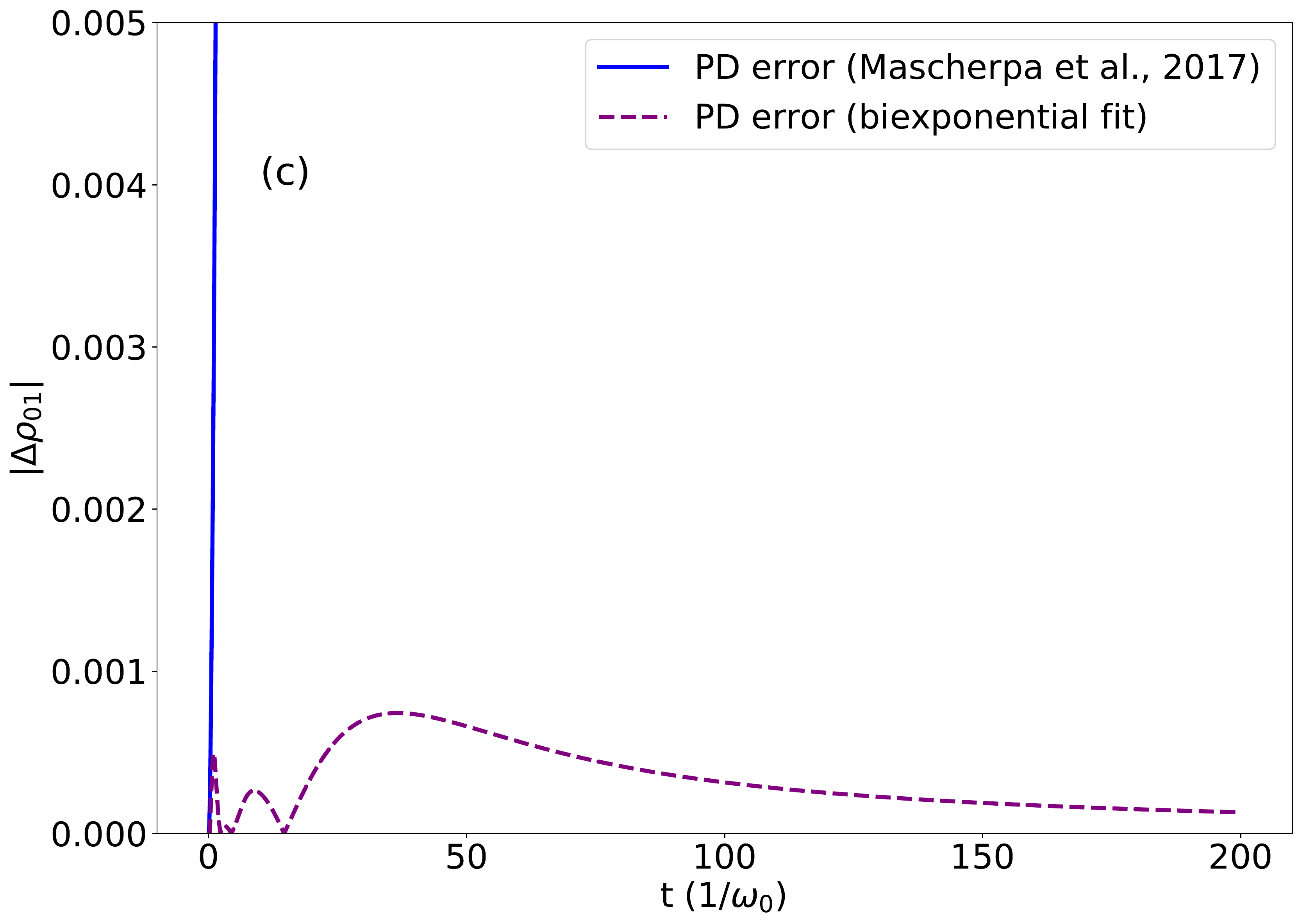}
\caption{\label{fig:errorplenio}Error in the dynamics given by the coherence $\rho_{01}$ term of the density matrix in the $\sigma_z$ basis by considering a pure dephasing model. We compare the error according to  \cite{PlenioErrorsCorr} against the error due to our Matsubara fitting approach. The error due to the fitting is computed by simulating the dynamics exactly by taking the full infinite Matsubara integral and then by considering only two terms from the fitting and finding the difference in the dynamics. The actual dynamics is \bc shown in the inset. \ec In the left figures, $\lambda = 0.2\omega_0$, $\gamma = 0.05\omega_0$, $\omega_q = 0$, $\Delta=0$, $T=0$. In the center figures  $\lambda = 0.4\omega_0$, $\gamma = 0.4\omega_0$.  In the right figures  $\lambda = \omega_0$, $\gamma = \omega_0$.  We see that in all cases, at long times, the dynamics is very sensitive to the error. In the very broad-bath case (c)  the performance in comparing to the pure dephasing results is misleading since the suppression of the error is just due to the very fast decay of the coherences.}
\end{figure*}
\end{widetext}

\subsection{Pure dephasing model}
The pure dephasing case is given by the condition $\Delta=0$ in the Hamiltonian in \eqref{spinh}. Since the pure dephasing case has an analytical solution, we can in principle also use it as a benchmark for comparing errors. The evolution of the density matrix is given, in the $\sigma_z$ basis, by \cite{gowan10},
\beq
\rho =
  \begin{bmatrix}
\rho_{00}(0) & \rho_{01}(0)e^{-F(t)} \nonumber \\
\rho_{10}(0)e^{-\bar F(t)} & \rho_{11}(0)
  \end{bmatrix}\;,
\eeq
with $F(t) = i\omega_q t + \int_0^t d\tau D(\tau)$, and $D(\tau)$ is defined as,
\beq
\label{dephasing_integral}
D(\tau) = 2 \int_0^{\tau} ds \left[ C(\tau - s) + \bar C(\tau - s) \right]\;,
\eeq
where $C(t)$ is the correlation function. \par

Let us write $C(t) = \sum_k c_k e^{\mu_kt}$, where $c_k$ and $\mu_k$ can be real or imaginary (note here that $c_k$ and $\mu_k$ refer to a generic decomposition of the correlation functions, not the one we define in the main text). This  easily allows us now to write $D(\tau)$ as a sum of exponentials as well. After integrating we again obtain a sum of exponentials,
\beq
I(\tau)  &=& \int_0^{\tau}dsC(\tau - s) = \int_0^{\tau}ds \sum_k c_k e^{\mu_k(\tau -s)} \nonumber \\
&=& \sum_k c_k \left [\frac{e^{\mu_k \tau}-1}{\mu_k}\right]\;\;.
\eeq
Using this expression into Eq.~(\ref{dephasing_integral}), we can write
\begin{eqnarray}
D(\tau) = 2 \sum_k \frac{c_k}{\mu_k}(e^{\mu_k \tau}-1)+\text{H.c.}\;
\end{eqnarray}
This gives
\begin{eqnarray}
\int_0^{t}d\tau D(\tau) = 2 \sum_k\left[\frac{c_k}{\mu_k^2}(e^{\mu_k t}-1)- \frac{c_k}{\mu_k}  t\right]+\text{H.c.}\;
\end{eqnarray}
Now, for any correlation function which is a sum of exponentials we can easily write down the \bc evolution as two parts: \ec  the sum of exponents from the non-Matsubara part and an integral taking into account the full Matsubara contribution, $\int_0^{t} D(\tau) = \int_0^{t} D_0(\tau) + \int_0^{t} D_\text{m}(\tau)$. In our case, the Matsubara terms are already given as an infinite sum of exponentials and in the zero temperature limit this is easy to calculate,
\begin{widetext}
\begin{equation}
{\renewcommand{\arraystretch}{2.3}
\begin{array}{lll}
\displaystyle\int_0^{t} dt D_\text{m}(\tau) &=& \displaystyle 4 \int_0 ^{t} d\tau \int_0^{\tau} ds \left[ - \frac{4 \lambda^2 \gamma}{\pi} \left(\frac{\pi}{\beta}\right)^2
\sum_{n =1}^{ \infty}\frac{ne^{-{2n\pi s}/{\beta} }}{[(\Omega + i \Gamma)^2 + ({2n\pi}/{\beta})^2][(\Omega - i \Gamma)^2 + ({2n\pi}/{\beta})^2]} \right] \nonumber \\
&=&\displaystyle -\frac{4\lambda^2 \gamma}{\pi} \int_0^{t} d\tau \int_0^{\tau} ds  \int_0^{\infty} dx\; \frac{x e^{-xs}}{[(\Omega + i \Gamma)^2 + x^2][(\Omega - i \Gamma)^2 + x^2]} \nonumber \\
&=&\displaystyle -\frac{4\lambda^2 \gamma}{\pi} \int_0^{\infty} dx \;\frac{1}{[(\Omega + i \Gamma)^2 + x^2][(\Omega - i \Gamma)^2 + x^2]} \left( t + \frac{e^{-xt}-1}{x} \right)\;\; \nonumber
\end{array}}
\end{equation}
\end{widetext}
where we took  $2n\pi/\beta\rightarrow x$. We use this expression to compare the dynamics of the pure dephasing model for the full Matsubara contribution against our approximation using just two exponents. In Figure (\ref{fig:errorplenio}) we show the comparison for different parameter regimes, and the bound in the same system quantities given by the inequality in \eqref{plenioineq}.

\bc Unfortunately, it becomes apparent from the figure that both the bound proposed in \cite{PlenioErrorsCorr}, and the pure dephasing result, are exponentially sensitive to errors in the fit at long times (when the error is comparable to the evolution time), which in the main text is one of the regimes we are interested in. However, it turns out that in terms of the influence of an error in the correlation functions on the system dynamics, the pure dephasing case is the worst case, as discussed \cite{PlenioErrorsCorr}, and hence, unfortunately, these results do not give us much information about the potential error in results away from the regime $\Delta=0$. In addition, for long time scales the error bound from  \cite{PlenioErrorsCorr} is a very weak bound.  A potential alternative method to characterize stability and error of results is discussed in the next section.\ec

\begin{figure}[h]
\includegraphics[width = 25em]{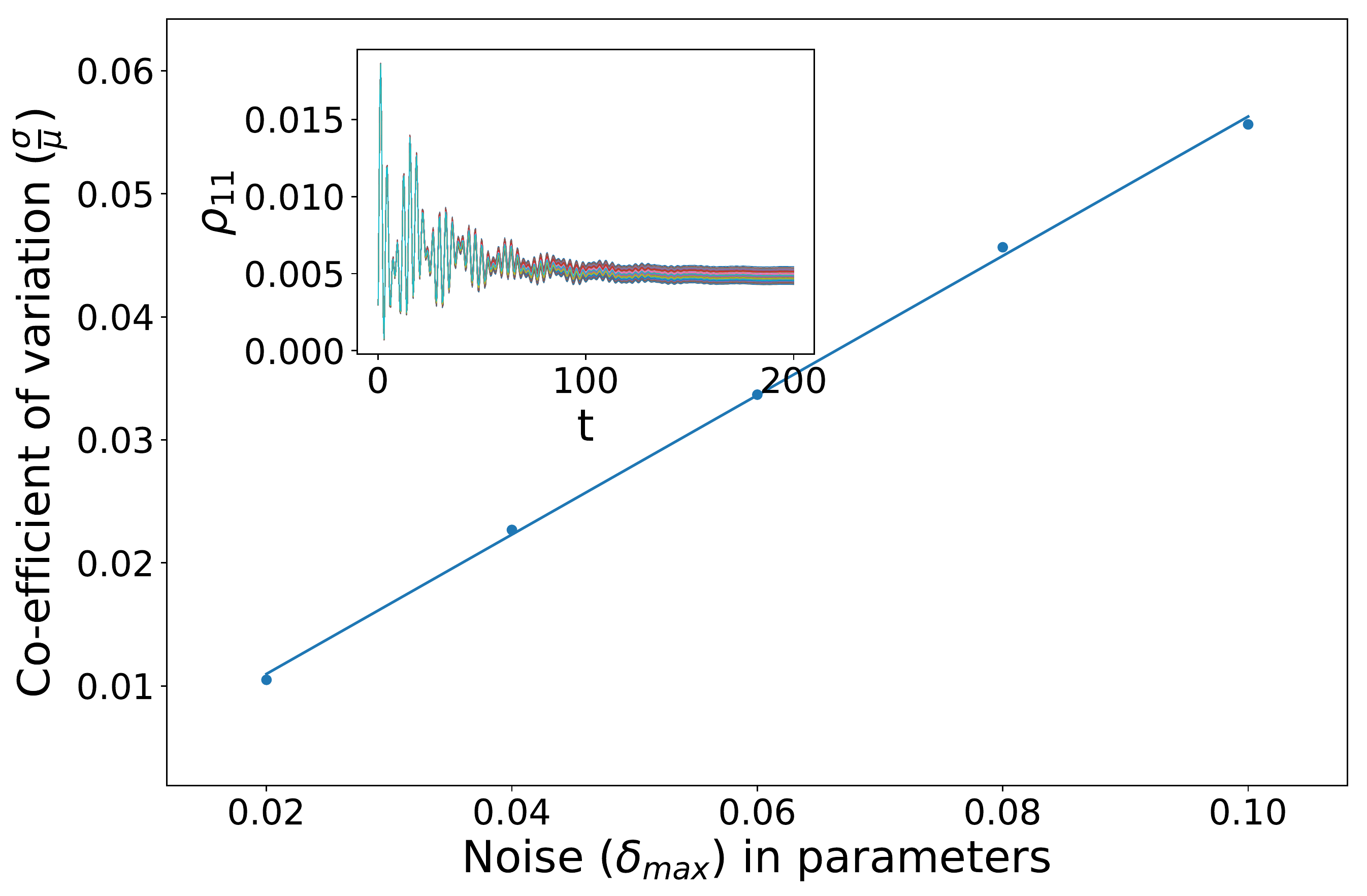}
\includegraphics[width = 25em]{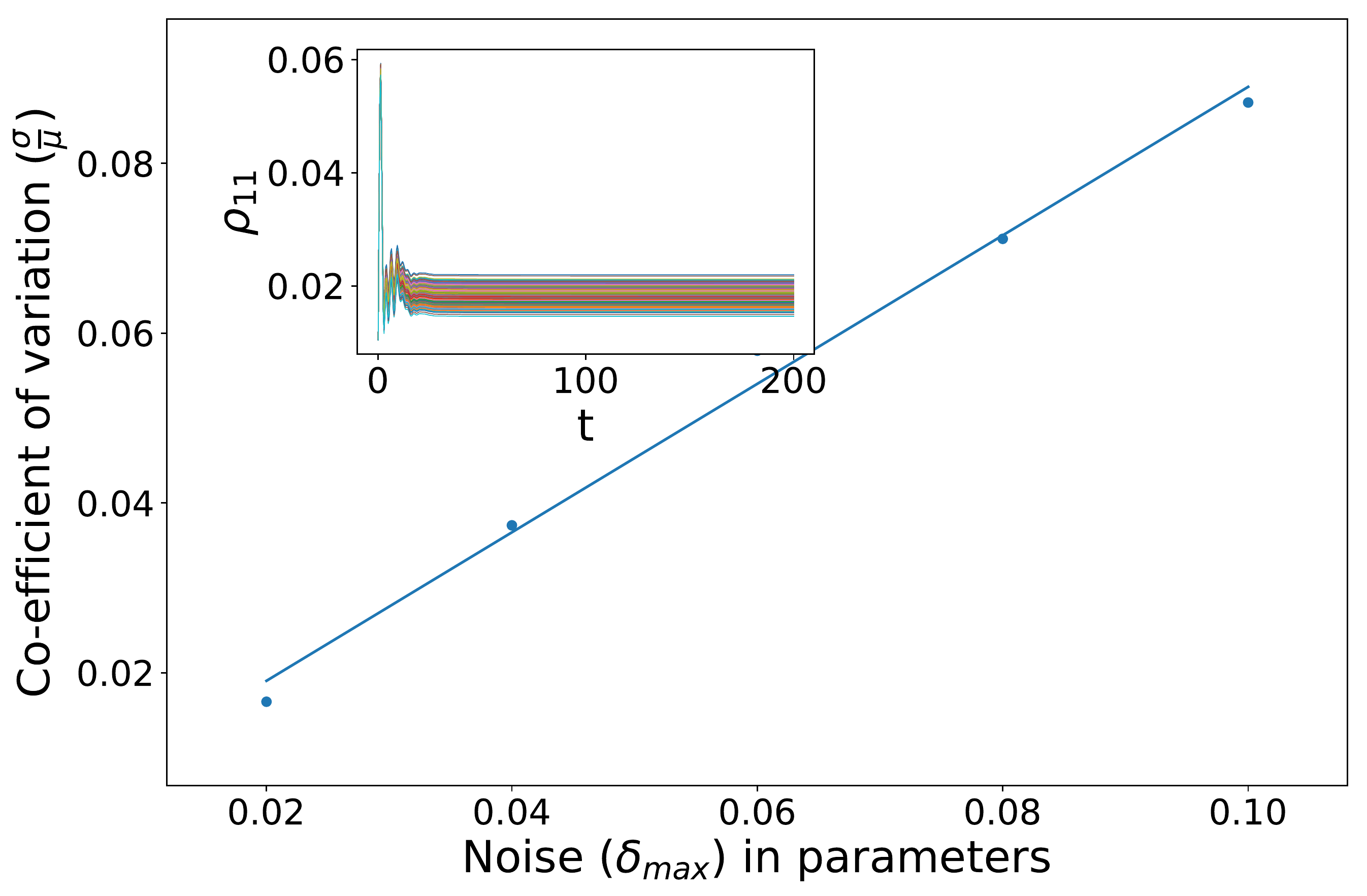}
\includegraphics[width = 25em]{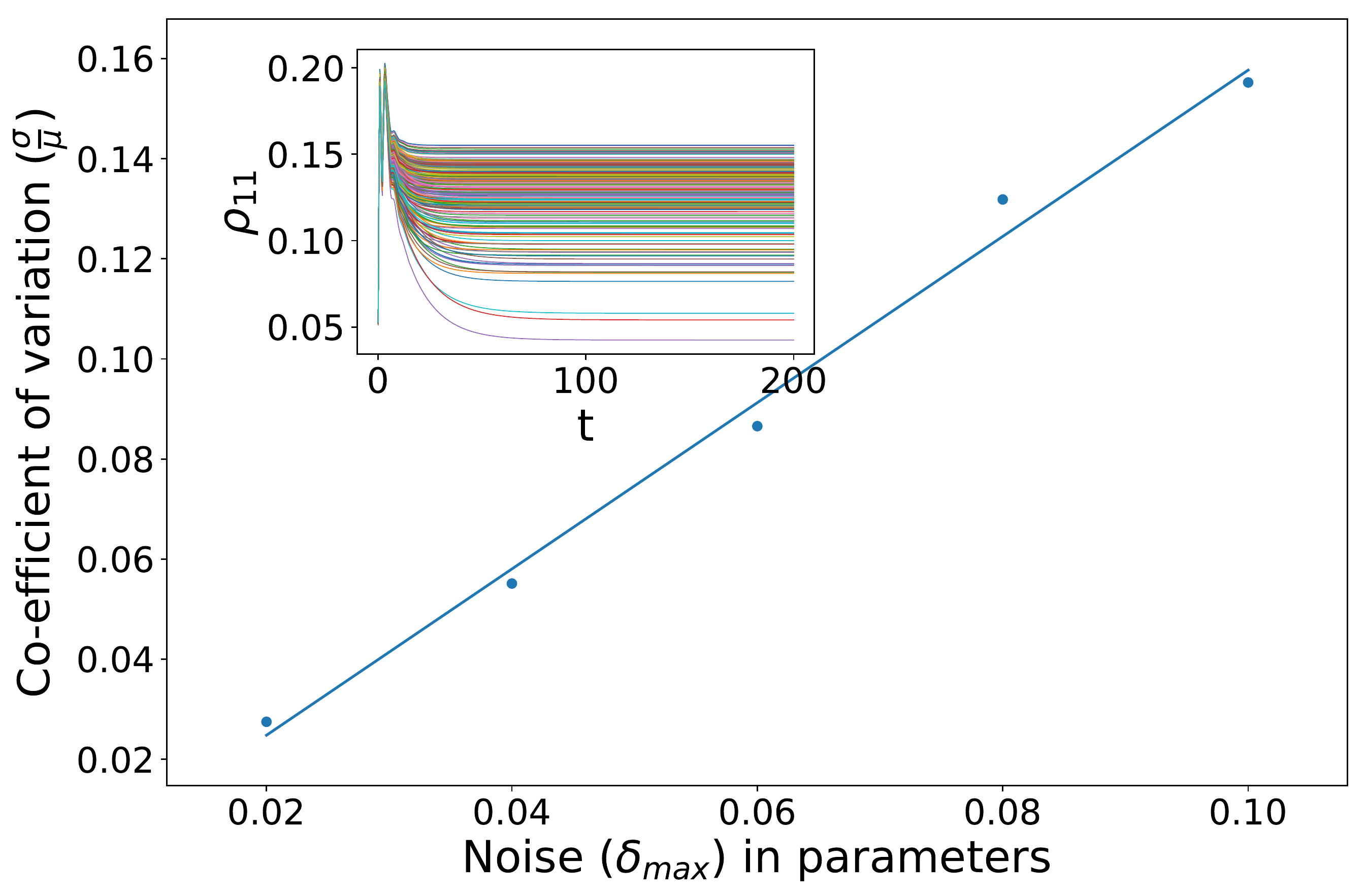}
\caption{\label{fig:sensitivity}The coefficient of variation, $\frac{\sigma}{\mu}$, of the steady state excited state population of the qubit against injected perturbations in the parameters of the biexponential fitting ($\pm \delta_{\mathrm{max}}$). In the top figure $\lambda = 0.2\omega_0$, $\gamma = 0.05\omega_0$, $\omega_q = 0$, $\Delta=\omega_0$, $T=0$. In the center figures  $\lambda = 0.4\omega_0$, $\gamma = 0.4\omega_0$.  In the bottom figures  $\lambda = \omega_0$, $\gamma = \omega_0$.  These results are averaged over 200 random choices of perturbed parameters. In the insets we show examples of the dynamics for perturbations upto 10\% in the parameters. As the perturbations decrease, we get better results and less deviation for the steady state populations.  }
\end{figure}

\section{Sensitivity of the dynamics to perturbations}

In order to further evaluate the sensitivity of the \bc dynamics and steady-state to the quality of the fitting of the Matsubara terms for $\Delta \neq 0$, \ec we numerically compute the evolution with \bc small random perturbations \ec added to the fit parameters. We use the standardized measure of dispersion of a distribution, the coefficient of variation, to quantify how much the steady-state population varies as we inject random perturbations to the parameters of our fitting. \par  The coefficient of variation is defined as the ratio between the standard deviation and mean (${\sigma}/{\mu}$) of observations. In this case, the observations are the steady-state populations of the system density matrix. The parameters that we will perturb are the amplitudes of the biexponentials ($c_1, c_2$) and the frequencies ($\mu_1, \mu_2$) in \eqref{biexp}. We inject perturbations as follows,
\beq
c_i \rightarrow c_i(1 + \delta) \nonumber\\
\mu_i \rightarrow \mu_i(1 + \delta), \nonumber
\eeq
where $\delta \in [-\delta_\text{max}, \delta_{\max}]$ is the perturbation in the parameters with maximum absolute value $\delta_\text{max}$. In Fig.~(\ref{fig:sensitivity}), we plot $({\sigma}/{\mu})$ against randomly picked values of $\delta$ from a uniform distribution and then compute the statistics after 200 runs.
\par
The intuition here is that these results show that additional small perturbations (errors) in the fitting parameters do not give a large variance in the results. Given that we also know the error in the fit without these additional perturbations, these results give us an intuition about how that error influences the steady-state of the system (see the next section for an example). Primarily however, these results show that as the perturbations/errors are decreased, the coefficient of variation for the steady state populations also decreases, suggesting that we can place a qualitative error bound on the final results.
\par

\section{Steady-state vs coupling}

\begin{figure}[h]
\includegraphics[width = 25em]{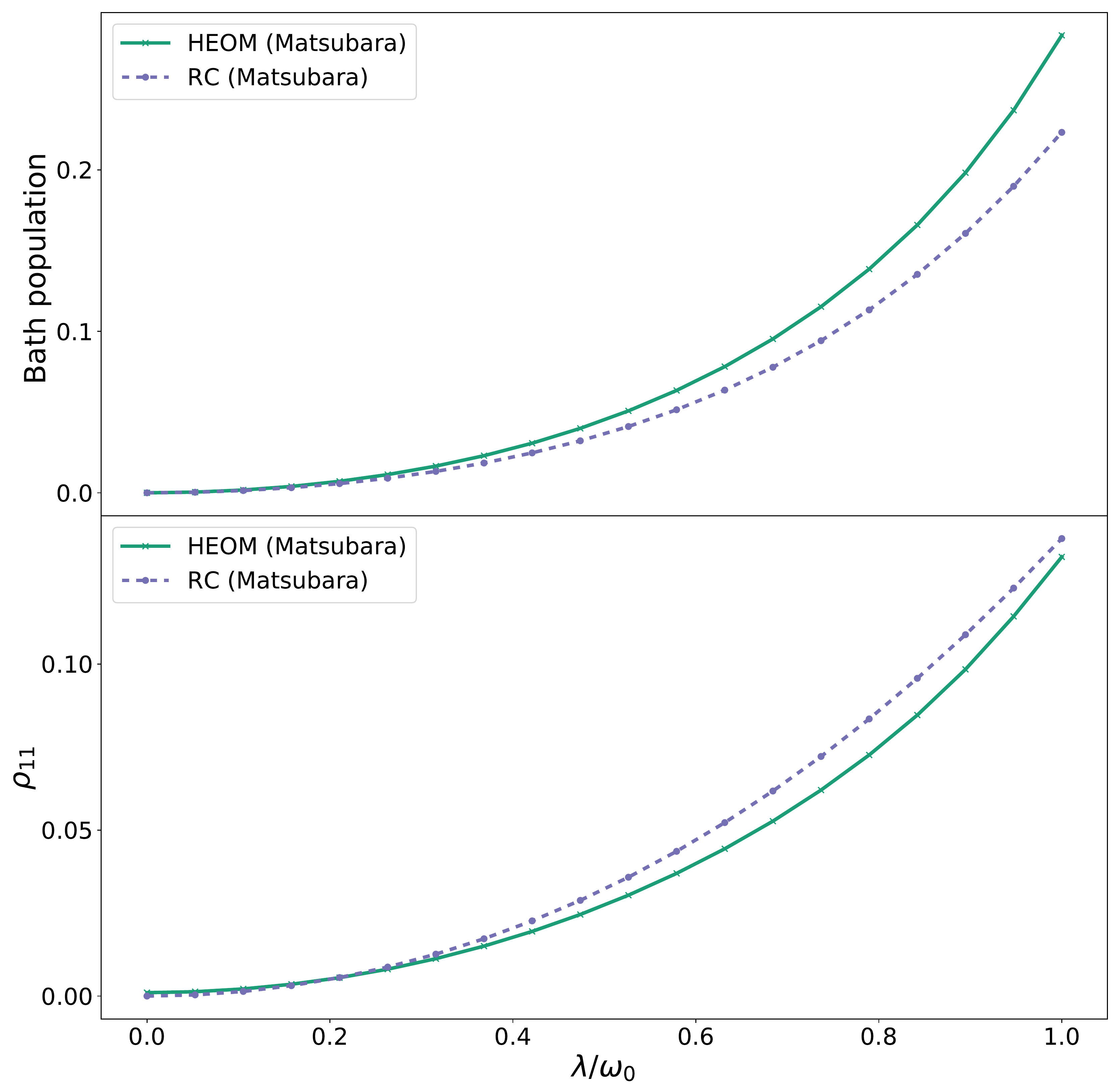}
\caption{\label{fig:steadystatepop} In the upper figure we plot the in the steady state population of the relevant effective ``bath mode'' against the coupling strength $\lambda$,  for $\omega_q = 0$, $\Delta=\omega_0$, $T=0$, and $\gamma=\omega_0$. In the lower figure we plot the qubit excited state probability for the same parameters. }
\end{figure}

As discussed in the main text, as the coupling strength increases, the HEOM and pseudomode predictions diverge from that of the RC model. \bc We also note that, as the coupling increases, the Matsubara terms become more important. 
\ec To clarify this, and give an example for the error analysis performed in the previous section, we compare the steady-state system excitation probability, and the bath-mode photon population, as a function of the coupling strength at zero temperature, see Figure (\ref{fig:steadystatepop}).\par
In the absence of a strong result on the error bounds of the various methods, we will try to make a qualitative argument here regarding the difference in the RC, HEOM/pseudomode predictions. Our sensitivity analysis in the previous section suggests that potential perturbations, or errors, in the fitting of the Matsubara terms can lead to errors in the steady-state population. From a direct comparison between the fit we use in this data, we estimate the parameter error in the fit to be about $~1\%$.  As we see from \figref{fig:sensitivity} an additional injected error of 1--2\% introduces a variance in the results at most 2--4\%, even in the USC regime. However, in figure \figref{fig:steadystatepop} we see that the difference in the RC versus HEOM/pseudomode results are much larger than this potential error from the inaccuracy in the fit \bc (especially in the broad-bath case, see \figref{fig:comparision-all}) \ec.

Thus, it would be reasonable to believe that this difference is not just an artifact of a poor fitting of the Matsubara terms but comes more from the RC approach being fundamentally inadequate in capturing the full correlations between the qubit and its environment for broad baths and strong couplings.  In addition, this reasoning suggests the fitting procedure we employ here can give reliable predictions upto a potential error of 2--4\% in the populations  in the long-time limit for the most difficult parameter choices (broad baths and strong couplings).

One other interesting error-related point in \figref{fig:steadystatepop} is the fact that the system population does not go perfectly to zero as $\lambda \rightarrow 0$ (the smallest value of $\lambda$ actually used in this figure is $1\times 10^{-5} \omega_0$).  This is because, as we saw in \figref{4b}, at weak coupling we still need the Matsubara terms to give correct detailed balance.  An equivalent plot without the Matsubara terms results in a residual excited state population of $\rho_{11}\approx0.055$, for $\lambda=  1\times 10^{-5} \omega_0$, whereas with the Matsubara terms included, that population extracted from the HEOM solution is $~0.001$.  In principle this small residual ``effective temperature'' is another indication of the quality of the fit, at least for small coupling strengths.

\section{Pseudomodes} \label{APP:PM}
In seminal work \cite{Garroway}, Garraway introduced the idea of modelling the dynamics of an open quantum system by replacing the environment with a set of bosonic pseudomodes. This can simplify the original problem in two ways. First, the infinite environmental degrees of freedom in the original system can be replaced by a finite set of modes. Second, the time-evolution of the pseudomodes can be captured by a Lindblad master equation.  However, in his examples, Garraway restricted himself to a rotating-wave-approximation form for the interaction between system and environment, and single excitations. Recently, his proof was formally extended by  Tamascelli \emph{et al.} \cite{plenio2018} to allow for non-RWA interactions.  However, here we need to adapt their proof to deal with the problem we face in our main text;  what happens if the correlation functions are negative?

In this section, we adapt  the results in \cite{plenio2018} to explicitly write a pseudomodes-model valid when the correlations of the original (Gaussian) bath can be written as a weighted-sum of exponentials. We show that when some of these weights are negative, the exact  system dynamics corresponds to a pseudomode model involving a modified quantum-mechanical equation of motion with a {\em non}-Hermitian Hamiltonian. Since approximating the Matsubara correlations in our main text with exponentials requires negative weights, this result has particular relevance in terms of restoring the correct non-Markovian and equilibrium physics.

After modelling the correlation function of the original spin-boson model as a sum of $N$ exponentials, we proceed in three steps. First, we map the  system dynamics to the situation in which the spin interacts with $N$ independent harmonic baths. Importantly, these baths follow a non-standard equation of motion when their Hamiltonian is non-Hermitian. Second, we show that each of these baths can be replaced by a non-Hermitian open quantum system involving a single pseudomode. The spectral density characterizing the interaction between each pseudomode and its residual environment is found to be constant for all positive and negative frequencies. Third, we show that this open quantum system is equivalent to imposing a pseudo-Schr\"{o}dinger master equation for each  pseudomode.\\

We stress that the steps above extend the work done in \cite{plenio2018} by explicitly showing that those results have physical meaning even when the initial exponential correlations have negative weight. In addition, at several crucial points, we restrict ourselves to the zero-temperature case, for notational simplicity.

\subsection{From one bath to $N$ baths}
 To set the notation, as in the main text we consider a system $S$ interacting with an environment $B$ of bosonic modes under the Hamiltonian
\begin{equation}
\label{eq:HHH}
H = H_S+ H_B+ \sigma_z \tilde{X}\;\;,
\end{equation}
where $b_k$ is the annihilation operators of the $k^{\text{th}}$ bath mode with energy $\omega_k$, and the interaction operator is $\tilde{X}=\sum_k \tilde{X}_k$, where  $\tilde{X}_k=g_k/\sqrt{2\omega_k}(b_k+b^\dagger_k)$. The Hamiltonian of the system and bath can be chosen to be $H_S= {\omega_q}/{2}~\sigma_z + {\Delta}/{2} ~ \sigma_x $ and $H_B=\sum_k \omega_k b_k^{\dagger}b_k $, respectively, as in Eq.~(\ref{spinh}). Importantly, we assume the initial state to be factorized as $\rho_S(0)\otimes \rho_B(0)$, where $\rho_S(0)$ is the initial state of the system, and where $\rho_B(0)$ is a Gaussian state of the bath satisfying $\text{Tr}_B[\tilde{X}\rho_B(0)]=0$.
The reduced evolution of the system $\rho_S(t)=\text{Tr}_B[\rho(t)]$ can be written as
\begin{widetext}
\begin{equation}
\label{eq:Dyson}
\begin{array}{lll}
\rho_S(t)&=&\displaystyle \sum_{n=0}^\infty(-i)^n\int_0^t d t_1\cdots\int_0^{t_{n-1}}d t_n\sum_{{n'=0}}^\infty(i)^{n'}\int_0^t d t'_{1}\cdots\int_0^{t'_{n'-1}}d t'_{n'}\\
&&\displaystyle\text{Tr}_B\left(\tilde{X}(t_1)\cdots\tilde{X}(t_n)\rho_B(0)\tilde{X}(t'_{n'})\cdots\tilde{X}(t'_1)\right)U_0(t)\sigma_z(t_1)\cdots\sigma_z(t_n)\rho_S(0)\sigma_z(t'_{n'})\cdots\sigma_z(t'_1)U_0^\dagger(t)\;\;,
\end{array}
\end{equation}
\end{widetext}
where $\tilde{X}(t)=\exp(iH_B t)\tilde{X}\exp(-i H_B t)$,  and $\sigma_z(t)=U_0^\dagger(t)\sigma_zU_0(t)$, with $U_0(t)=\exp(-i H_S t)$.
Since the initial state of the bath is Gaussian and such that $\text{Tr}_B[\tilde{X}\rho_B(0)]=0$, the correlations $\text{Tr}_B\left(\tilde{X}(t_1)\cdots\tilde{X}(t_n)\rho_B(0)\tilde{X}(t'_{n'})\cdots\tilde{X}(t'_1)\right)$ appearing in the equation above can, in principle, be retrieved from the two-time correlation
\begin{equation}
\label{eq:corr_app}
C(t)=\text{Tr}_B[\tilde{X}(t)\tilde{X}(0)]\;\;.
\end{equation}
For this reason, the  reduced Dyson equations in Eq.~(\ref{eq:Dyson}) is invariant under splitting of the original bath $B$ into $N$ independent copies $B_i$ (with initial Gaussian state $\rho_{B_i}(0)$) described by the total Hamiltonian
\begin{equation}
H' = H_S+ \sum_{i=1}^{N} H'_{B_i} + \sigma_z\sum_{i=1}^{N} \tilde{X}_i\;\;,
\end{equation}
and such that the two-time correlation functions are constrained by
\begin{equation}
\label{eq:CorrNbaths}
\text{Tr}_{B_1}\cdots\text{Tr}_{B_N}\left[\left(\prod_{i=1}^{N}\rho_{B_i}(0)\right)\sum_{i=1}^{N}\tilde{X}_i(t)\sum_{j=1}^{N}\tilde{X}_j(0)\right]=C(t),
\end{equation}
where $C(t)$ is the original correlation function in Eq.~(\ref{eq:corr_app}). In the equations above, $H'_{B_i}$ and $\tilde{X}_i$ are the free-bath Hamiltonian and coupling operator with support on the bath $B_i$. Note that, as before, the time dependence in Eq.~(\ref{eq:CorrNbaths}) follows the free-bath Hamiltonian $\tilde{X}_i(t)=\exp(i H'_{B_i})\tilde{X}_i\exp(-i H'_{B_i} t)$. Since the baths are independent, the constraint in Eq.~(\ref{eq:CorrNbaths}) can be written as
\begin{equation}
\begin{array}{lll}
C(t)&=&\displaystyle\sum_{i=1}^{N}\text{Tr}[\rho_{B_i}(0)\tilde{X}_i(t)\tilde{X}_i(0)]\\
&&+\displaystyle\sum_{i\neq j}\text{Tr}_{B_i}\left[\rho_{B_i}(0)\tilde{X}_i(t)\right]\text{Tr}_{B_j}\left[\rho_{B_i}(0)\tilde{X}_j(0)\right].
\end{array}
\end{equation}
To satisfy the equation above it is sufficient to impose
\begin{equation}
\label{eq:conditions}
\begin{array}{llll}
\displaystyle\sum_{i=1}^{N}\text{Tr}_{B_i} [\rho_{B_i}(0)\tilde{X}_i(t) \tilde{X}_i(0)]&=&C(t)&\\
\displaystyle\text{Tr}_{B_i}\left(\rho_{B_i}(0)\tilde{X}_i(0)\right)&=&0&\forall i=1,\cdots,N\;\;.
\end{array}
\end{equation}
The simplicity of decomposing the original bath into $N$ independent ones as just described hides an important point. In fact, since we are only interested in the  dynamics of the reduced system $\rho_S(t)$, we can let the coupling operators  $\tilde{X}_i$ (and hence $H'$) to be non-Hermitian, as long as they satisfy the contraints in Eq.~(\ref{eq:conditions}) \emph{and} give rise to equations of motion in the same form as in  Eq.~(\ref{eq:Dyson}) with the substitution $\tilde{X}\mapsto\sum_{i=1}^N\tilde{X}_i$. To ensure the latter, we  need to \emph{impose} the equation of motion
\begin{equation}
\label{eq:Hermitian}
\frac{d}{dt}\rho'(t)=-i[H',\rho'(t)]\;\;.
\end{equation}
We here explicitly stress that, for a non-Hermitian Hamiltonian $H'$ the usual Shr\"{o}dinger dynamics would imply the right-hand side of the previous equation to take the form$-i[H'\rho'(t)-\rho(t){H'}^\dagger]$. Here however,  in order to ensure the invariance of the Dyson equation, we need to impose Eq.~(\ref{eq:Hermitian}) instead. Under these hypothesis
\begin{equation}
\rho_S'(t)=\rho_S(t)\;\;,
\end{equation}
where $\rho'_S=\text{Tr}_{B_1}\cdots\text{Tr}_{B_n}[\rho'(t)]$.\\

\subsection{From $N$ baths to $N$ pseudomodes}
Following \cite{plenio2018}, we now can proceed a step further to show that each of the baths $B_i$ can be replaced by a single pseudomode (associated to a Hilbert space $R_i$, annihilation operator $a_i$, and frequency $\Omega_i$) interacting with a residual environment $E_i$ (whose modes are associated with annihilation operators $b_{i,\alpha}$ and have frequency $\omega_{i,\alpha}$) so that the full now Hamiltonian reads
\begin{equation}
\label{eq:newH}
\begin{array}{lll}
H''&=&\displaystyle H_S+H''_B+\sigma_z\sum_{i=1}^{N} \tilde{X}^a_i
\end{array}
\end{equation}
where $\tilde{X}^a_i=\lambda_i/\sqrt{2\Omega_i}(a^\dagger_i+a_i)$, with the parameters $\lambda_i$ setting the scale for the interaction between the pseudomodes and the system.  We also defined the free-bath Hamiltonian as
\begin{equation}
\label{eq:HRE}
H''_B=\sum_{i=1}^{N} H''_{B_i}\;\;,
\end{equation}
where
\begin{equation}
\begin{array}{lll}
H_{B_i}''&=& \displaystyle\Omega_ia^\dagger_i a_i+i\sum_{\alpha}\frac{g_{i,\alpha}}{\sqrt{2\omega_{i,\alpha}}}(b^\dagger_{i,\alpha} a_{i}-a^\dagger_i b_{i,\alpha})\\
&&+\displaystyle\sum_{\alpha}\omega_{i,\alpha}b^\dagger_{i,\alpha}b_{i,\alpha}\;\;.
\end{array}
\end{equation}
The interaction of each pseudomode with its residual environment $E_i$ is described by the parameters $g_{i,\alpha}$ which, in the continuum limit, are charaterized by the spectral densities
\begin{equation}
\label{eq:JJJ}
\begin{array}{lll}
J_{i}(\omega)&=&\displaystyle\pi\sum_{\alpha}\frac{g^2_{i,\alpha}}{2\omega_{i,\alpha}}\delta(\omega-\omega_{i,\alpha})\;\;.
\end{array}
\end{equation}
We now \emph{impose} the pseudo-equation of motion (see Eq.~(\ref{eq:Hermitian}))
\begin{equation}
\label{eq:eq:mot_11}
\frac{d}{dt}\rho''(t)=-i[H'',\rho''(t)]\;\;,
\end{equation}
where we stress again that, since the Hamiltonian $H''$ can, in principle, be non-Hermitian, these equations might be non-standard. Analogously to Eq.~(\ref{eq:conditions}), we also impose the following constraints on the correlations
\begin{equation}
\label{eq:corrA}
\begin{array}{lll}
\displaystyle\sum_i\text{Tr}_{R_i}\text{Tr}_{E_i} [\rho_{R_i}(0)\rho_{E_i}(0)\tilde{X}^a_i(t) \tilde{X}^a_i(0)]&=&C(t)\\
\displaystyle\text{Tr}_{R_i}\text{Tr}_{E_i}\left(\rho_{R_i}(0)\rho_{E_i}(0)\tilde{X}^a_i(0)\right)&=&0\;\;,
\end{array}
\end{equation}
for an initial environmental state of the form $\prod_i[\rho_{R_i}(0)\rho_{E_i}(0)]$, where $\rho_{R_i}(0)$ and $\rho_{E_i}(0)$ are the initial Gaussian state of the $i^\text{th}$ pseudomode and its residual environment, respectively. In the expression above, the time evolution follows $\tilde{X}^a_i(t)=\exp(i H''_{B_i}t)\tilde{X}^a_i \exp(-iH''_{B_i}t)$.
Following the same considerations as above, on the equivalence between two open quantum systems, Eq.~(\ref{eq:eq:mot_11}) and Eq.~(\ref{eq:corrA}) are sufficient for the reduced dynamics $\rho_S''(t)=\text{Tr}_{R}\text{Tr}_{E}[\rho''(t)]$ (where $R=\prod_i R_i$ and $E=\prod_i E_i$) to exactly match the original one, i.e.,
\begin{equation}
\label{eq:restwoapix}
\rho_S''(t)=\rho_S'(t)=\rho_S(t)\;\;.
\end{equation}

From Eq.~(\ref{eq:corrA}), we see that the correlation $C(t)$ effectively induces constraints on the spectral densities in Eq.~(\ref{eq:JJJ}) and the couplings $\lambda_i$. Specifically,  choosing $J_i(\omega)$ to be constant for both positive and negative frequencies, i.e.,
\begin{equation}
\label{eq:SD_app}
\begin{array}{lllll}
J_i(\omega)&=&\displaystyle\frac{\gamma_i}{2}\;\;,
\end{array}
\end{equation}
and assuming all the pseudomodes and their residual environments  to be initially in their ground state, the reduced dynamics of the system is the same as that of the original spin-boson model with correlations
\begin{equation}
\label{eq:Corr_temp}
C(t)=\frac{\lambda_i^2}{2\Omega_i}\sum_{i=1}^N e^{-\displaystyle (i\Omega_i + \gamma_i/2)t}\;\;.
\end{equation}
To show this, we solve the Heisenberg equation of motion for the free bath and insert the result in Eq.~(\ref{eq:corrA}). We start by noticing that the equal-time commutation relations $[b_{i,\alpha}(t),b^\dagger_{j,\beta}(t)]=\delta_{i j}\delta_{\alpha\beta}$, $[a_i(t),a^\dagger_j(t)]=\delta_{ij}$, and $[b_{i,\alpha}(t),a_{i}(t)]=[b_{i,\alpha}(t),a^\dagger_{i}(t)]=0$ are satisfied once we impose them as an initial condition (the dynamics of the open quantum system is unitary). We can now formally write the equations of motion for the residual environments $E_i$ as
\begin{equation}
\frac{d}{dt}b_{i,\alpha}=i[H_{B}'',b_{i,\alpha}]=-i\omega_{i,\alpha}b_{i,\alpha}+\frac{g_{i,\alpha}}{2\sqrt{\omega_{i,\alpha}}}a_i\;\;,
\end{equation}
which leads to the following equations for the corresponding Laplace transforms (denoted by an overhead bar)
\begin{equation}
\label{eq:temp_1}
\begin{array}{lll}
\bar{b}^\dagger_{i,\alpha}+\bar{b}_{i,\alpha}&=&\displaystyle\left(\frac{b^\dagger_{i,\alpha}(0)}{s-i\omega_{i,\alpha}}+\frac{b_{i,\alpha}(0)}{s+i\omega_{i,\alpha}}\right)\\
&&+\displaystyle\frac{g_{i,\alpha}\left[s(\bar{a}^\dagger_i+\bar{a}_i)+i\omega_{i,\alpha}(\bar{a}^\dagger_i-\bar{a}_i)\right]}{\sqrt{2\omega_{i,\alpha}}(s^2+\omega^2_{i,\alpha})}\\
\bar{b}^\dagger_{i,\alpha}-\bar{b}_{i,\alpha}&=&\displaystyle\left(\frac{b^\dagger_{i,\alpha}(0)}{s-i\omega_{i,\alpha}}-\frac{b_{i,\alpha}(0)}{s+i\omega_{i,\alpha}}\right)\\
&&+\displaystyle\frac{g_{i,\alpha}\left[s(\bar{a}^\dagger_i-\bar{a}_i)+i\omega_{i,\alpha}(\bar{a}^\dagger_i+\bar{a}_i)\right]}{\sqrt{2\omega_{i,\alpha}}(s^2+\omega^2_{i,\alpha})}\;\;,
\end{array}
\end{equation}
where $s$ is the complex variable introduced by the Laplace transformation.
Similarly, we can write the Heisenberg equations for the pseudomodes as
\begin{equation}
\dot{a}_i=i[H_{B}'',a_{i}]=-i\Omega_ia_i-\sum_\alpha \frac{g_{i,\alpha}}{\sqrt{2\omega_{i,\alpha}}}b_{i,\alpha}\;\;,
\end{equation}
which, after a Laplace transform, reads
\begin{equation}
\label{eq:temp_2}
\begin{array}{lll}
s \bar{x}_i &=&\displaystyle x_i(0)+\Omega_i\bar{p}_i-\sum_{\alpha}\frac{g_{i,\alpha}}{\sqrt{2\omega_{i,\alpha}}}(\bar{b}_{i,\alpha}^\dagger+\bar{b}_{i,\alpha})\\
s \bar{p}_i &=&\displaystyle p_i(0)-\Omega_i \bar{x}_i-i\sum_{\alpha}\frac{g_{i,\alpha}}{\sqrt{2\omega_{i,\alpha}}}(\bar{b}_{i,\alpha}^\dagger-\bar{b}_{i,\alpha})\;\;,
\end{array}
\end{equation}
in terms of the dimensionless quadratures $x_i=a^\dagger_i+a_i$ and $p_i=i(a^\dagger_i-a_i)$. By inserting Eq.~(\ref{eq:temp_1}) into Eq.~(\ref{eq:temp_2}) we finally obtain
\begin{equation}
\label{eq:temp_LEQ}
\begin{array}{lll}
s \bar{x}_i &=&\displaystyle x_i(0)+\left(\Omega_i-\sum_\alpha\frac{g^2_{i,\alpha}}{2(s^2+\omega^2_{i,\alpha})}\right) \bar{p}_i\\
&&\displaystyle -s\sum_{\alpha}\frac{g^2_{i,\alpha}}{2\omega_{i,\alpha}(s^2+\omega^2_{i,\alpha})}\bar{x}_i-{x}_i^\text{in}\\
s \bar{p}_i &=&\displaystyle p_i(0)-\left(\Omega_i-\sum_\alpha\frac{g^2_{i,\alpha}}{2(s^2+\omega^2_{i,\alpha})}\right) \bar{x}_i\\
&&\displaystyle -s\sum_{\alpha}\frac{g^2_{i,\alpha}}{2\omega_{i,\alpha}(s^2+\omega^2_{i,\alpha})}\bar{p}_i-{p}_i^\text{in}\;\;,
\end{array}
\end{equation}
where
\begin{equation}
\begin{array}{lll}
{x}_i^\text{in}&=&\displaystyle\sum_{i,\alpha}\frac{g_{i,\alpha}}{\sqrt{2\omega_{i,\alpha}}}\left(\frac{b^\dagger_{i,\alpha}(0)}{s-i\omega_{i,\alpha}}+\frac{b_{i,\alpha}(0)}{s+i\omega_{i,\alpha}}\right)\\
{p}_i^\text{in}&=&\displaystyle i\sum_{i,\alpha}\frac{g_{i,\alpha}}{\sqrt{2\omega_{i,\alpha}}}\left(\frac{b^\dagger_{i,\alpha}(0)}{s-i\omega_{i,\alpha}}-\frac{b_{i,\alpha}(0)}{s+i\omega_{i,\alpha}}\right)\;\;.
\end{array}
\end{equation}
Using Eq.~(\ref{eq:JJJ}), we can write Eq.~(\ref{eq:temp_LEQ}) in the continuum limit as
\begin{equation}
\label{eq:tempppp_aaaa}
\begin{array}{lll}
s \bar{x}_i &=&\displaystyle x_i(0)+ \left[\Omega_i-\int_{-\infty}^{\infty} d\omega\frac{ J_i(\omega)\omega}{\pi(s^2+\omega^2)}\right] \bar{p}_i\\
&&\displaystyle -s\int_{-\infty}^\infty\frac{J_i(\omega)}{\pi(s^2+\omega^2)}\bar{x}_i-{x}_i^\text{in}\\
s \bar{p}_i &=&p_i(0)-\displaystyle \left[\Omega_i-\int_{-\infty}^{\infty} d\omega\frac{J_i(\omega)\omega}{\pi(s^2+\omega^2)}\right] \bar{x}_i\\
&&\displaystyle -s\int_{-\infty}^\infty\frac{J_i(\omega)}{\pi(s^2+\omega^2)}\bar{p}_i-{p}_i^\text{in}\;\;.
\end{array}
\end{equation}
By inserting Eq.~(\ref{eq:SD_app}) into the equation above we obtain
\begin{equation}
\label{eq:tempp}
\begin{array}{lll}
s \bar{x}_i &=&\displaystyle x_i(0) +\Omega_i\bar{p}_i-{\gamma_i}/{2}~\bar{x}_i-{x}_i^\text{in}\\
s \bar{p}_i &=&\displaystyle p_i(0)-\Omega_i\bar{x}_i-{\gamma_i}/{2}~\bar{p}_i-{p}_i^\text{in}\;\;.
\end{array}
\end{equation}
Note that, from Eq.~(\ref{eq:HRE}), it seems that we have not introduced the correct renormalization terms for the frequency of the pseudomodes. In fact, this is justified by the choice of spectral densities in Eq.~(\ref{eq:SD_app}) as the additional term which renormalizes the frequencies in Eq.~(\ref{eq:tempppp_aaaa}) is  $\displaystyle\lim_{\Lambda\rightarrow\infty}\int_{-\Lambda}^{\Lambda} d\omega\frac{ J_i(\omega)\omega}{\pi(s^2+\omega^2)}=0$, where we regularized the integral at infinity. For this reason, the frequencies $\Omega_i$ already correspond to the correctly renormalized ones.
Using the equation above we find the following equation for the Laplace transform of the quadratures $x_i$
\begin{equation}
[(s+\gamma_i/2)^2+\Omega^2_i]\bar{x}_i=(s+\gamma_i/2)[x_i(0)-{x}^\text{in}]+\Omega_i[p_i(0)-{p}^\text{in}_i].
\end{equation}
which we can insert into the first of Eq.~(\ref{eq:corrA}) to finally obtain the correlation function as
\begin{equation}
\label{eq:corr_form}
\begin{array}{lll}
C(t)&=&\displaystyle\sum_{i=1}^{N}\frac{\lambda_i^2}{2\Omega_i}\mathcal{L}_t^{-1}\left\{\text{Tr}_{R_i}\text{Tr}_{E_i}[\rho_{R_i}(0)\rho_{E_i}(0)\bar{x}_ix_i(0)]\right\}\\
&=&\displaystyle\sum_{i=1}^{N}\frac{\lambda^2_i}{2\Omega_i}\frac{1}{2\pi i}\int ds\left\{\frac{[s+\gamma_i/2]\langle x(0)x(0)\rangle_{R_i}}{(s+\gamma_i/2)^2+\Omega^2_i}e^{st}\right.\\
&&+\displaystyle\left.\frac{\Omega_i\langle p(0)x(0)\rangle_{R_i}}{(s+\gamma_i/2)^2+\Omega^2_i}e^{st}\right\}\\
&=&\displaystyle\sum_{i=1}^{N}\frac{\lambda^2_i}{2\Omega_i}e^{-\displaystyle (i\Omega_i+\gamma_i/2)t}\;\;,
\end{array}
\end{equation}
where  $\tilde{X}^a_i=\lambda_i/\sqrt{2\Omega_i}x_i$ and we defined $\langle\cdot\rangle_{R_i}\equiv\text{Tr}_{R_i}[\;\cdot\;\rho_{R_i}(0)]$ as the trace over the $i^\text{th}$ pseudomode, and  $\mathcal{L}^{-1}_t$ as the inverse Laplace transform. We also used
 $\langle x_i(0)\rangle_{R_i}=0$,  together with $\langle x_i(0) x_i(0)\rangle_{R_i}=1$ and $\langle p_i(0) x_i(0)\rangle_{R_i}=-i$, since $\rho_{R_i}(0)$ is the pseudomodes' ground-state. This correlation is the same as the one in Eq.~(\ref{eq:Corr_temp}) which is the result we wanted to prove to deduce Eq.~(\ref{eq:restwoapix}).

\subsection{The pseudo-Lindblad equation}

Following \cite{plenio2018,Gardiner04}, we can now complete the third step promised at the beginning of this section, i.e., showing that the reduced dynamics of the system can be obtained by considering the following effective pseudo-Lindblad dynamics for the system and the pseudomodes $R_i$
\begin{equation}
\label{eq:Lpm}
\frac{d}{dt}\rho_\text{pm}= L[\rho_\text{pm}]\;\;,
\end{equation}
where $L[\rho]=-i[H_\text{pm},\rho]+\sum_i D_i[\rho]$,  where the pseudomodes Hamiltonian reads
\begin{equation}
H_\text{pm}=H_S+\sigma_z\sum_i\tilde{X}^a_i+\sum_{i=1}^{N}\Omega_i a^\dagger_i a_i\;\;,
\end{equation}
 and where $D_i[\rho]=\gamma_i/2\left[2a_i\rho a_i^\dagger-(a_i^\dagger a_i\rho+\rho a_i^\dagger a_i)\right]$. As before, when $H_\text{pm}$ is non-Hermitian, the equation of motion above are non-standard.

To proceed in the proof, we use Eq.~(\ref{eq:Lpm}) to find that all operators $\hat{O}_{SR}(t)$ with support on the system and pseudomodes space satisfy the equation of motion
\begin{equation}
\label{eq:tt_11}
\frac{d}{dt}\langle[\hat{O}_{SR}]\rangle_{SR}=i\langle[H_\text{pm},\hat{O}_{SR}]\rangle_{SR}+\sum_{i=1}^{N} \langle[D_i^\dagger(\hat{O}_{SR})]\rangle_{SR}\;\;,
\end{equation}
where we defined $\langle\cdot\rangle_{SR}=\text{Tr}_{SR}[\;\cdot\;\rho_{SR}(0)]$ [with $\rho_{SR}(0)=\rho_S\prod_{i=1}^N\rho_{R_i}(0)$] and where $D_i^\dagger[\cdot]$ is the adjoint of the operator $D_i[\cdot]$ with respect to the trace, i.e., $\text{Tr}_{SR}[D_i(\hat{A})\hat{B}]=\text{Tr}_{SR}[\hat{A}D_i^\dagger(\hat{B})]$ for any operator $\hat{A}$, $\hat{B}$ with support on the system and pseudomode Hilbert spaces $SR$. Specifically, $D^\dagger_i[\rho]=\gamma_i/2\left[2a_i^\dagger\rho a_i-(a_i^\dagger a_i\rho+\rho a_i^\dagger a_i)\right]$.
In parallel, from Eq.~(\ref{eq:newH}), we obtain the following  Heisenberg equation of motion
\beq
\label{eq:temp_aa}
\displaystyle\frac{d}{dt}\langle[\hat{O}_{SR}]\rangle_{SRE}&=&i\displaystyle\langle[H'',\hat{O}_{SR}]\rangle_{SRE}\\
&=&i\displaystyle\langle[H_\text{pm},\hat{O}_{SR}]\rangle_{SRE}\nonumber\\
&&-\displaystyle \sum_{i,\alpha}\frac{g_{i,\alpha}}{\sqrt{2\omega_{i,\alpha}}}\langle b^\dagger_{i,\alpha}[a_i,\hat{O}_{SR}]\rangle_{SRE}\nonumber\\
&&+\displaystyle \sum_{i,\alpha}\frac{g_{i,\alpha}}{\sqrt{2\omega_{i,\alpha}}}\langle[a^\dagger_i,\hat{O}_{SR}] b_{i,\alpha}\rangle_{SRE}\;\;,\nonumber
\eeq
where $\langle\cdot\rangle_{SRE}=\text{Tr}_{E_1}\cdots\text{Tr}_{E_N}\text{Tr}_{SR}$. To close the equations above, we need to compute the equation of motion $\displaystyle\frac{d}{dt}b_{i,\alpha}=i[H'',b_{i,\alpha}]$ for the operators $b_{i,\alpha}(t)$ of the residual baths. This leads to a result which is equivalent to Eq.~(\ref{eq:temp_1}), and which reads
\begin{equation}
\label{eq:temptemptemp_00}
\displaystyle\sum_{\alpha}\frac{g_{i,\alpha}}{\sqrt{2\omega_{i,\alpha}}}\bar{b}_{i,\alpha}(t)=b^\text{in}_{i}(t)+A_i(t)\;\;,
\end{equation}
where
\beq
b^\text{in}_{i}(t)&=&\displaystyle\mathcal{L}_t^{-1}\left[\sum_{\alpha}\frac{g_{i,\alpha}{b}_{i,\alpha}(0)}{\sqrt{2\omega_{i,\alpha}}(s+i\omega_{i,\alpha})}\right]\\
A_i(t)&=&\displaystyle\mathcal{L}_t^{-1}\left[\sum_{\alpha}\frac{g^2_{i,\alpha}\bar{a}_{i}}{{2\omega_{i,\alpha}}(s+i\omega_{i,\alpha})}\right]\;\;.\nonumber
\eeq
Now, using the convolution theorem and  the identities $\mathcal{L}_t^{-1}\left[{1}/{(s+i\omega)}\right]=e^{-i\omega t}$ and $\mathcal{L}_t^{-1}\left[\bar{a}_i\right]=a_i(t)$, we notice that, in the continuum limit, the last term in the previous expression can be written as
\beq
\label{eq:temptemptemp_11}
A_i(t)&=&\displaystyle\frac{1}{\pi}\mathcal{L}_t^{-1}\left[\int_{-\infty}^\infty d\omega \frac{J_i(\omega)}{s+i\omega}\bar{a}_i\right]\nonumber\\
&=&\displaystyle \frac{\gamma_i}{2\pi}\int_{-\infty}^\infty d\omega \int_0^t dt' a_i(t')e^{-i\omega (t-t')}\\
&=&\displaystyle\frac{\gamma_i}{2}a_i(t)\;\;,\nonumber
\eeq
where we used $\int_{-\infty}^\infty d\omega e^{i(t'-t)}=2\pi\delta(t-t')$ and $\int_0^t dt'\delta(t-t)=1/2$ (see \cite{Gardiner04}, Eq. 5.3.12).\\
Using Eq.~(\ref{eq:temptemptemp_11}) and Eq.~(\ref{eq:temptemptemp_00}) into Eq.~(\ref{eq:temp_aa}), allows us to write
\begin{widetext}
\beq
\label{eq:tt_22}
\displaystyle\frac{d}{dt}\langle\hat{O}_{SR}(t)\rangle_{SRE}&=&i\displaystyle\langle[H_\text{pm},\hat{O}_{SR}(t)]\rangle_{SRE}-\displaystyle \sum_{i=1}^{N}\frac{\gamma_i}{2}\left[ \langle a^\dagger_i(t)[a_i(t),\hat{O}_{SR}(t)]\rangle_{SRE}- \langle [a^\dagger_i(t),\hat{O}_{SR}(t)]a_i(t)\rangle_{SRE}\right]\nonumber\\
&=&i\displaystyle\langle[H_\text{pm},\hat{O}_{SR}(t)]\rangle_{SRE}+\displaystyle \sum_{i=1}^{N}\langle D_i^\dagger[\hat{O}_{SR}(t)]\rangle_{SRE}\;\;,
\eeq
\end{widetext}
where we used the fact that, since the residual environment is in the ground state, $b_{i,\alpha}(0)\rho_{E_i}(0)=\rho_{E_i}(0) b^\dagger_{i,\alpha}(0)=0$. We can now notice that Eq.~(\ref{eq:tt_11}) and Eq.~(\ref{eq:tt_22}), despite referring to different underlying spaces, lead to the very same set of closed equation for operators with support in $SR$, hence predicting the same physical dynamics in such a space. In the Schr\"{o}dinger picture this results in
\begin{equation}
\text{Tr}_{R}[\rho_\text{pm}(t)]=\rho''_S(t)=\rho_S(t)\;\;,
\end{equation}
where we used Eq.~(\ref{eq:restwoapix}). This completes our proof.

For completeness, it is also interesting to explicitly show that Eq.~(\ref{eq:Lpm})  gives, indeed, the same correlations as in Eq.~(\ref{eq:corr_form}). In particular, we want to compute the correlations for the ``free'' pseudomodes, i.e.,
\begin{equation}
\label{eq:Cpm}
C_\text{pm}(t)=\text{Tr}_{R}\left[F(t) F(0) \rho_R(0)\right]\;\;,
\end{equation}
where $F(t)=e^{{L}_R^\dagger t}[F(0)]$, with  $L_R[\;\cdot\;]=-i[\sum_i\Omega_i a^\dagger_i a_i,\;\cdot\;]+\sum_i D_i[\;\cdot\;]$, and where $F(0)=\sum_{i=1}^{N}\tilde{X}^a_i=\sum_{i=1}^{N} \lambda_i/\sqrt{2\Omega_i}(a^\dagger_i+a_i)$. We further defined $\rho_R(0)=\prod_i\rho_{R_i}(0)$, where  $\rho_{R_i}(0)$ is the initial state of each pseudomodes (which, as before, we assume to be the ground state). From its definition, we note that $F(t)=\sum_{i=1}^N\lambda_i/\sqrt{2\Omega_i}[a^\dagger_i(t)+a_i(t)]$ where $a^\dagger_i(t)+a_i(t)=e^{L^\dagger_R t}[a^\dagger_i+a_i]$. The operator $a^\dagger_i(t)+a_i(t)$ can be found solving the coupled differential equation (to be compared with Eq.~(\ref{eq:tempp})
\begin{equation}
\begin{array}{lll}
\displaystyle \frac{d}{dt}(a^\dagger_i(t) +a_i(t))&=&L^\dagger_R[a^\dagger_i(t) +a_i(t)]\\
&=&\displaystyle i \Omega_i (a^\dagger_i(t)-a_i(t))-\frac{\gamma_i}{2}(a^\dagger_i(t)+a_i(t))\\
\displaystyle\frac{d}{dt}(a^\dagger_i(t) -a_i(t))&=&L^\dagger_R[a^\dagger_i(t) -a_i(t)]\\
&=&\displaystyle i \Omega_i (a^\dagger_i(t)+a_i(t))-\frac{\gamma_i}{2}(a^\dagger_i(t)-a_i(t)),
\end{array}
\end{equation}
whose solution can be plugged into Eq.~(\ref{eq:Cpm}) to obtain
\begin{equation}
C_\text{pm}(t)=C(t)\;\;.
\end{equation}
\subsection{Modelling the absence of Matsubara correlations}
In this subsection we apply the previous analysis to the case in which the full correlation function in Eq.~(\ref{corrspecial}) is approximated  as
\begin{equation}
C(t)\rightarrow C_0(t)=\frac{\lambda^2}{2\Omega}e^{-i\Omega t}e^{-\gamma/2 t}\;\;,
\end{equation}
i.e., we completely neglect the Matsubara correlations in Eq.~(\ref{matsubaracorr}) in the zero temperature limit. From Eq.~(\ref{eq:HRE}) we find that this corresponds to an open quantum system in which a single pseudomode (with annihilation operator $a$) mediates the interaction between the system and the residual environment (with modes associated to annihilation operators $b_\alpha$ and frequency $\omega_\alpha$) as
\begin{equation}
\label{eq:Heff}
\begin{array}{lll}
H_{\text{Mats}} &=& \displaystyle H_S + \sigma_z \frac{\lambda}{\sqrt{ 2 \Omega }} (a + a^{\dagger}) + \Omega a^{\dagger}a\\
&&+ \displaystyle\sum_\alpha \omega_\alpha b_\alpha^{\dagger} b_\alpha +\sum_\alpha \frac{g_\alpha}{\sqrt{2\Omega} \sqrt{2\omega_\alpha }} \left(b_\alpha^{\dagger}a-a^\dagger b_\alpha\right).
\end{array}
\end{equation}
As described in Eq.~(\ref{eq:SD_app}), the coupling $g_\alpha$ to the residual environment are determined, in the continuum limit, by the spectral density
\begin{equation}
J_\text{Mats}(\omega)=\gamma\Omega\;\;.
\end{equation}
Note that the apparent additional $2\Omega$ factor in the equation above with respect to Eq.~(\ref{eq:SD_app}) just reflects a different definition of the residual couplings in Eq.~(\ref{eq:Heff}) with respect to Eq.~(\ref{eq:HRE}).
In alternative, from the results in the previous section, we also find that the system dyanmics can be found by solving
\begin{equation}
\label{eq:rhoMats}
\frac{d}{dt}\rho_{\text{eff}} = -i[H_\text{eff},\rho_{\text{Mats}}] + D_{\text{Mats}}[\rho_{\text{Mats}}]\;\;,
\end{equation}
where
\begin{equation}
\label{eq:temptemptemp}
\begin{array}{lll}
H_{\text{eff}} &=&\displaystyle  H_S + \sigma_z \frac{\lambda}{\sqrt{ 2 \Omega }} (a + a^{\dagger}) + \Omega a^{\dagger}a \\
D_{\text{Mats}}[\rho_{\text{Mats}}] &=& \displaystyle \frac{\gamma}{2}\left(2a \rho_{\text{Mats}} a^{\dagger} - a^{\dagger}a\rho_{\text{Mats}} - \rho_{\text{Mats}} a^{\dagger}a\right)\;\;,
\end{array}
\end{equation}
and tracing out the pseudomode from $\rho_{\text{Mats}}$.
The  equation of motion in Eq.~(\ref{eq:rhoMats}) describes the effect of neglecting the Matsubara correlations which are needed to model the correct  equilibrium and non-Markovian physics. Consistently, the Lindblad operator in Eq.~(\ref{eq:temptemptemp}) does not describe a residual bath at thermal equilibrium as it does not leave the eigenstates of the system-pseudomode Hamiltonian $H_\text{eff}$ invariant. This can lead to the possibility of peculiar effects such as ground state decay and, in gerenal, to a constant dissipation of energy in the steady state.
Interestingly, it has been shown (see Appendix A for a brief overview) that a modified version of the  model  in Eq.~(\ref{eq:Heff}) can be derived from mapping the original environment into a single  ``reaction coordinate'' and a residual (perturbative) environment.
The differences between the two models can be intuitively ascribed to performing a rotating-wave and Markov approximations (in the coupling with the residual bath). Within the perturbative limits for the coupling to the residual environment, the reaction-coordinate model leads to master equations \cite{iles2014environmental,Iles_Smith_2015,Strasberg_2016} which improve on Eq.~(\ref{eq:rhoMats}) to correctly describe the equilibrium and Markovian physics of the original spin-boson model.

\end{appendix}

\end{document}